**Correlated ionisations in two spatially separated nanometric volumes in the track structure of $^{241}$Am alpha particles: comparison with Monte Carlo simulations**

Gerhard Hilgers[1], Thomas Braunroth[1,2], Hans Rabus[1,3]

[1] Physikalisch-Technische Bundesanstalt (PTB), 38116 Braunschweig, Germany
[2] Present Address: Gesellschaft für Anlagen- und Reaktorsicherheit (GRS), 50667 Köln, Germany
[3] Present Address: Physikalisch-Technische Bundesanstalt (PTB), 10587 Berlin, Germany

**Abstract**

The production of two double strand breaks in spatially separated locations on the DNA molecule can cause the loss of a whole DNA loop, which can be of substantial length depending on the geometrical position of the two damaged sites and depends on the degree of correlation between ionisation clusters formed in sites of several nanometres in size. In the first part of this paper, nanodosimetric measurements of alpha particle tracks in 1.2 mbar $H_2O$, 1.2 mbar $C_3H_8$ and 1.2 mbar $C_4H_8O$ with the PTB ion counter nanodosimeter were reported. In this second part, the focus is on the geometrical characterisation of the two sites simulated with the nanodosimeter in the three target gases and on the comparison of the measurement results with Monte Carlo simulations. The measurements in 1.2 mbar $C_3H_8$ were simulated with a version of the track structure code PTra dedicated to modelling the PTB ion counter nanodosimeter. Further simulations were performed with Geant4-DNA for $^{241}$Am alpha particle tracks in liquid water. Simulations of the measurements and the actual measurement results are found to be in good agreement for the investigated irradiation geometries.

## 1. Introduction

The pattern of inelastic interactions in subcellular targets, especially in DNA, is substantially responsible for radiation-induced damage to tissue (Brenner and Ward, 1992; Goodhead, 2006). Therefore, besides absorbed dose, the track structure of ionising particles is particularly important for the biological effectiveness of ionising radiation (Grosswendt, 2005; Conte et al., 2017). Consequently, the understanding of the detailed spatial pattern of ionisations is a crucial factor for predicting radiation effects in biological tissue. The required spatial resolution for the local distribution of energy deposition corresponds to the size of the relevant biological entities. It ranges from a few micrometres (cell nucleus) down to the spacing of the DNA double helix of two nanometres. The induction of double strand breaks or even clustered damage to the DNA is considered the decisive mechanism for cell inactivation and cancer induction (Brenner and Ward, 1992; Goodhead, 2006; Grosswendt, 2005; Conte et al., 2017).

Although there are radiobiological data indicating the relevance of correlations in damages to DNA (Simmons and Watts, 1999; Friedrich et al., 2018), it has been traditionally assumed that the local ionisation density in a single nanometre-sized site (which is represented by the ionisation cluster size (ICS) distribution) is the most important quantity (Grosswendt, 2005; 2002). However, it can be expected that particularly for denser ionising particles also the spatial correlation of clusters becomes important. In recent attempts to develop a radiation action model based on nanodosimetry, the occurrence of ionisation clusters in spatial proximity is an essential ingredient of the model that was developed for sparsely ionising photon radiation (Schneider et al, 2020). Further analysis of the model foundations revealed the importance of correlated as opposed to coincident occurrence of ionization clusters (Ngcezu and Rabus, 2021).

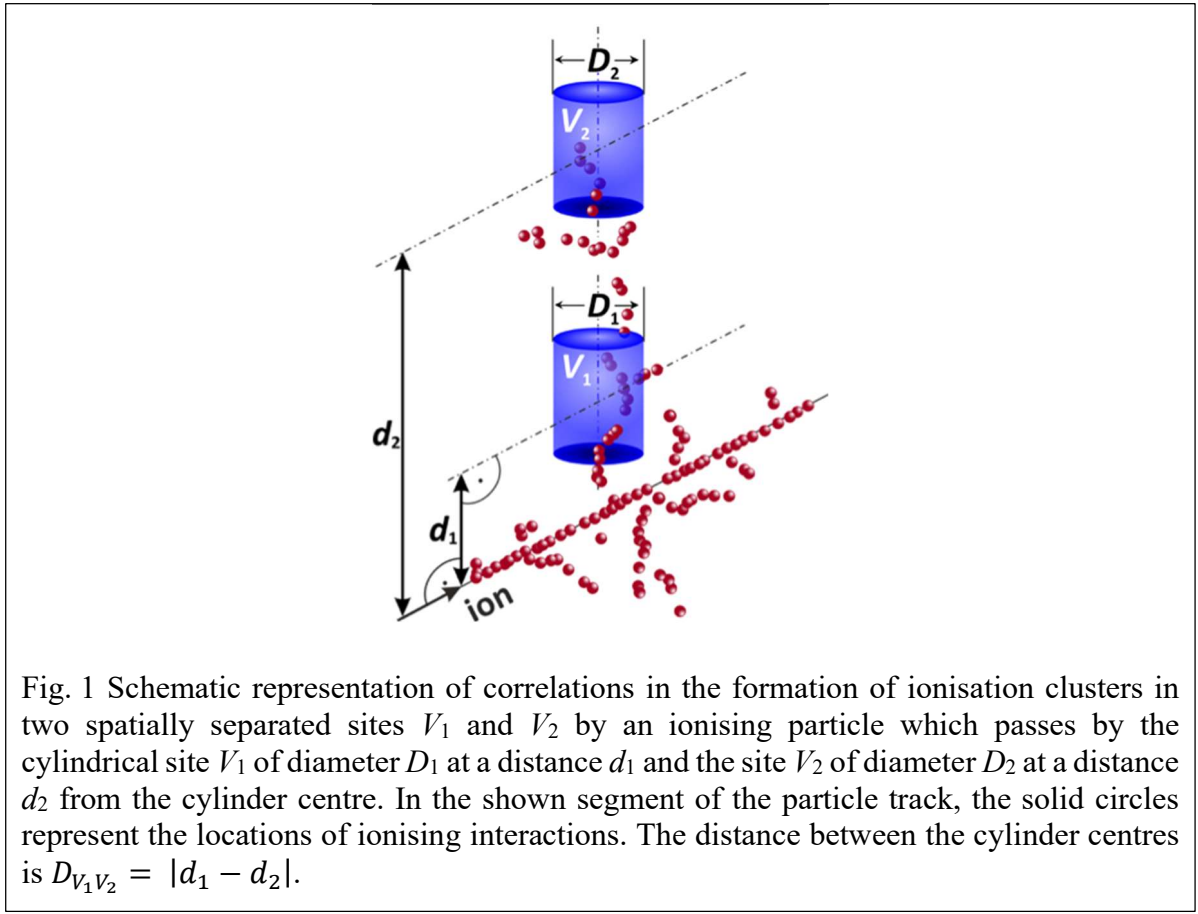

Fig. 1 Schematic representation of correlations in the formation of ionisation clusters in two spatially separated sites $V_1$ and $V_2$ by an ionising particle which passes by the cylindrical site $V_1$ of diameter $D_1$ at a distance $d_1$ and the site $V_2$ of diameter $D_2$ at a distance $d_2$ from the cylinder centre. In the shown segment of the particle track, the solid circles represent the locations of ionising interactions. The distance between the cylinder centres is $D_{V_1V_2} = |d_1 - d_2|$.

From the biological point of view this is evident, because an isolated DNA damage, even a double-strand break, generally can be repaired. However, the occurrence of double or even multiple clustered damages producing two double strand breaks in two spatially separated sites, each of several nanometres in size, can cause the loss of a whole DNA loop (Harder, 1988). This lost DNA loop can be of substantial length depending on the geometrical position of the two sites and the degree of damage produced in the respective sites by a passing primary particle. Clustered damages in close proximity are more abundant in the Bragg peak region of tracks from heavier charged particles as compared to electrons and photons.

In the first part of this paper (Hilgers and Rabus, 2020), measurements have been reported that were carried out using the PTB ion counter nanodosimeter operated with three different target gases, $H_2O$ (water vapor), $C_3H_8$ (propane) and $C_4H_8O$ (tetrahydrofuran). A two-dimensional position sensitive detector (PSD) was used for registration of the primary particles, and the time differences between detection of the primary particle and detection of each secondary ion were recorded. Application of appropriate drift time windows in the off-line data analysis allowed realisation of two spatially separated target volumes (sites), one stacked above the other as shown schematically in Fig.1. The targets were represented by the spatial distribution of the mean ICS $M_1(d,h)$ as a function of the coordinates of arrival of the primary particle at the PSD, where $d$ and $h$ are the horizontal and vertical deviation from the PSD centre. In this part of the paper, the outcomes of Monte Carlo simulations of the experiment and of idealised target geometries are compared with each other and with the measured data to determine the effective (nanometric) target size and the distance between the target sites.

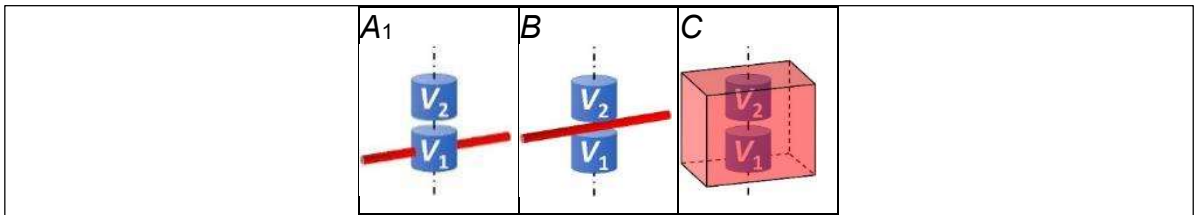

Fig. 2 Illustration of the three irradiation geometries considered for comparison of simulations and experiments. $A_1$: beam through centre of site $V_1$, $B$: beam through the centre point between the centres of $V_1$ and $V_2$, $C$: beam extending over an area that completely covers both sites $V_1$ and $V_2$ (broad beam irradiation).

## 2. Material and Methods

A crucial characteristic of the sites is their effective site size, defined as the size of a target cylinder producing ICS distributions (ICSDs) identical to those measured in the sites, if the probability for counting an ionisation inside the cylinder is unity and zero outside. In the first part of this paper (Hilgers and Rabus, 2020), the effective sizes of the sites $V_1$ and $V_2$ in the experiments were determined from the spatial distribution of the measured mean ICS and the spatial distribution of the simulated extraction efficiency. Furthermore, the effective target sizes in the measurements were scaled to equivalent effective site sizes in liquid $H_2O$ at unit density using the procedure described in (Grosswendt, 2006).

### 2.1. Scaling of the spatial distribution of the extraction efficiency in terms of liquid $H_2O$ and track structure simulations

In order to obtain spatial distributions of the extraction efficiency in terms of liquid $H_2O$, all three linear spatial dimensions of the extraction efficiency were scaled according to the scaling procedure described in (Grosswendt, 2006) using the ($\rho\lambda_{ion}$)-ratios applied in (Hilgers and Rabus, 2020; Hilgers et al., 2021), with $\lambda_{ion}$ denoting the mean free path for ionisation of the alpha particles in the respective materials calculated from the known ionisation cross sections and $\rho$ denoting the density. Further details on the determination of the spatial distribution of the extraction efficiency are discussed in (Hilgers et al., 2021).

In the present work, an approach based on track structure simulations using the PTra Monte Carlo code (Grosswendt, 2002; Bug et al., 2013) and Geant4-DNA, version 10.04.p02 (Incerti et al., 2010a; 2010b; Bernal et al., 2015; Incerti et al., 2018) with standard option "opt2" for electron transport was used to determine the site sizes and distances between the sites in the measurements and in terms of liquid $H_2O$, respectively, by applying the procedure presented in (Hilgers et al., 2021). As cross sections for the other two gases were not implemented in the PTra code, the simulations of the measurements could only be performed for $C_3H_8$.

With both codes, two types of simulation were performed. In the first type, the spatial distribution of the extraction efficiency was used in a Russian-roulette approach to decide whether a secondary ion produced by interactions was detected. In the second approach, the targets were cylinders for which the probability for counting an ionisation was unity when it occurred inside the cylinder and zero if it occurred outside. Further details on the track structure simulations (primary particle type and energy, beam size and profile) are discussed in (Hilgers et al., 2021).

In the PTra $C_3H_8$ track structure code, the scoring of ionisations was implemented in the code and performed during the track simulations. For Geant4-DNA, the alpha particle tracks were simulated in a slab of liquid $H_2O$ of thickness 270 nm, and all ionisations in a layer between 196 nm and 212 nm were recorded together with their respective coordinates to allow for superposition of particle track structure and target geometry. The simulation results were then processed with a dedicated program, in which the particle tracks were superimposed with the simulated spatial distribution of the extraction geometry for the respective target gas scaled in terms of liquid $H_2O$ or with the extraction efficiency for target cylinders stacked above each other to determine the ICSD.

The considered irradiation geometries were that the primary particle beam traverses either $V_1$ or $V_2$ centrally (only the former case $A_1$ is shown in Fig. 2), that it passes through the centre point between the centres of $V_1$ and $V_2$ (case $B$ in Fig. 2) or that it covers both targets (broad beam geometry $C$ in Fig. 2).

## 2.2. Optimisation procedure and used metrics

Scoring of the simulated tracks using the simple geometrical shape of two stacked cylinders was performed for a range of values of the parameters describing the cylindrical volume, $D_{eff}$ and $H_{eff}$, and the distance between the centres of the two stacked cylinders, $D_{V_1V_2}$. The parameter value(s) that gave the best agreement between measurement and simulation were determined by minimisation of two different metrics for the deviation between measured and simulated ICSDs.

Two different approaches were used to determine the parameters of the simple geometry of two stacked cylinders. In the first approach the parameters of the geometry were optimised consecutively. The two metrics applied in this approach are the same as in (Hilgers et al., 2021), i.e. the reduced $\chi^2$ and $\kappa$, which corresponds to the test quantity for a Kolmogorov-Smirnov test. In the first step the parameters describing the cylindrical volume, $D_{eff}$ and $H_{eff}$, were determined using the arithmetic mean of the two marginal distributions of primary alpha particles passing the target cylinders $V_1$ and $V_2$ centrally (geometry $A_1$ and $A_2$). Additionally, a proportional relation $H_{eff} = F \cdot D_{eff}$ was assumed with the ratio $F = H_{eff}/D_{eff}$ being identical to that resulting from $D_{eff}$ and $H_{eff}$ obtained by integration of the extraction efficiency as described in section 2.3 in (Hilgers et al., 2021). After optimising $D_{eff}$ (and, with it, $H_{eff}$), in the second step a similar procedure was carried out to optimise $D_{V_1V_2}$ keeping the previously optimised $D_{eff}$ and $H_{eff}$. Here the simulations were performed for a single cylinder with the alpha particles passing "below" the bottom of the cylinder (PTra) or for a pair of congruent cylinders with irradiation geometry $B$ in Fig. 2 (Geant4-DNA), and the experimental reference data was the arithmetic mean of the marginal distributions in the two targets for the case that the alpha particle beam passed centrally between the two targets.

In the calculation of both metrics for marginal ICSDs only those data $P_\nu^{EXP}(G)$ and $P_\nu^{MC}(G)$ were included for which the complementary cumulative probability

$$F_{v_c+1}^{EXP}(G) = 1 - \sum_{v=0}^{v_c} P_v^{EXP}(G) > 5 \cdot 10^{-5} \quad \text{with: } v_c \leq v_{max} \quad (1)$$

In principle, the data $P_v^{EXP}(G)$ and $P_v^{MC}(G)$ for which the complementary cumulative probability is larger than $(1 - 5 \cdot 10^{-5})$ should also be excluded. However, in all ICSDs $P_0^{EXP}(G)$ and $P_0^{MC}(G)$ exceeded $5 \cdot 10^{-5}$.

To account for the differences of the sites $V_1$ and $V_2$ in the measurements, the mean $P_v^{EXP}(G)$

$$P_v^{EXP}(A) = \frac{N_v^{EXP}(V_1, A_1) + N_v^{EXP}(V_2, A_2)}{N_{tot}^{EXP}(V_1, A_1) + N_{tot}^{EXP}(V_2, A_2)} \quad \text{for geometry } A \quad (2)$$

$$P_v^{EXP}(B) = \frac{N_v^{EXP}(V_1, B) + N_v^{EXP}(V_2, B)}{N_{tot}^{EXP}(V_1, B) + N_{tot}^{EXP}(V_2, B)} \quad \text{for geometry } B \quad (3)$$

$$\text{with: } \quad N_{tot}^{EXP}(V, G) = \sum_{v=0}^{v_{max}} N_v^{EXP}(V, G) \quad (4)$$

of the distributions measured in geometry $B$ and in the equivalent geometries $A_1$ and $A_2$ was used for comparison with the simulations. In eqs. (2) to (4), $N_v^{EXP}(V_1, G)$ and $N_v^{EXP}(V_2, G)$ are the number of ionisation clusters of size $v$ measured in $V_1$ and $V_2$, respectively, and $N_{tot}^{EXP}(V_1, G) = N_{tot}^{EXP}(V_2, G)$ are the total number of ionisation clusters measured in $V_1$ ($V_2$) in geometry $G$.

In the second approach the parameters $D_{eff}$, $H_{eff}$ and $D_{V_1V_2}$ describing the geometry were optimised simultaneously. This approach was applied only to simulations using Geant4-DNA, since the present version of PTra does not allow the simulation of two stacked target cylinders. As before, a proportional relation $H_{eff} = F \cdot D_{eff}$ was assumed with the ratio $F = H_{eff}/D_{eff}$ being identical to that resulting from $D_{eff}$ and $H_{eff}$ obtained by integration of the spatial distribution of the extraction efficiency. In contrast to the first approach, which considered the one-dimensional marginal distributions, this approach takes into account the entire bivariate distributions. Consequently, also the metrices required in the optimisation must be able to process two-dimensional data sets. Again, the first metric was the reduced $\chi^2$, i.e. $\chi^2(G) / n$, which in two dimensions was calculated according to:

$$\frac{\chi^2(G)}{n} = \frac{1}{n} \sum_{\mu=0}^{\mu_{max}} \sum_{v=0}^{v_{max}} \frac{\left(P_{\mu,v}^{EXP}(G) - P_{\mu,v}^{MC}(G)\right)^2}{\left(u\left(P_{\mu,v}^{EXP}(G)\right)\right)^2 + \left(u\left(P_{\mu,v}^{MC}(G)\right)\right)^2} \quad (5)$$

where $P_{\mu,v}^{EXP}(G)$ and $P_{\mu,v}^{MC}(G)$ are the measured and simulated data, respectively. $G$ denotes the set of parameters describing the geometry settings. The sum includes only those summands, where $P_{\mu,v}^{EXP}(G) > 0$ or $P_{\mu,v}^{MC}(G) > 0$, and $n$ is the number of data points for which $P_{\mu,v}^{EXP}(G) > 0$ or $P_{\mu,v}^{MC}(G) > 0$. $u(P_{\mu,v}^{EXP}(G))$ and $u(P_{\mu,v}^{MC}(G))$ denote the statistical uncertainties of $P_{\mu,v}^{EXP}(G)$ and $P_{\mu,v}^{MC}(G)$, respectively, which are defined by:

$$u\left(P_{\mu,\nu}(G)\right) = \sqrt{\frac{P_{\mu,\nu}(G) \cdot \left(1 - P_{\mu,\nu}(G)\right)}{N_{tot}(G)}} \quad \text{with:} \quad N_{tot}(G) = \sum_{\mu=0}^{\mu_{max}} \sum_{\nu=0}^{\nu_{max}} N_{\mu,\nu}(G) \quad (6)$$

$N_{\mu,\nu}(G)$ denotes the absolute frequency of ionisation cluster pairs with cluster sizes $\mu$ and $\nu$ for the first and the second cluster.

The second metric $\kappa(G)$ was extended to two dimensions according to the scheme presented in (Peacock, 1983; Fasano and Franceschini, 1987).

To account for the differences of the sites $V_1$ and $V_2$ in the measurements, again the mean $P_{\mu,\nu}^{EXP}(G)$

$$P_{\mu,\nu}^{EXP}(A) = \frac{N_{\mu,\nu}^{EXP}(V_1, V_2, A_1) + N_{\nu,\mu}^{EXP}(V_2, V_1, A_2)}{N_{tot}^{EXP}(V_1, V_2, A_1) + N_{tot}^{EXP}(V_2, V_1, A_2)} \quad \text{for geometry } A \quad (7)$$

$$P_{\mu,\nu}^{EXP}(G) = \frac{N_{\mu,\nu}^{EXP}(V_1, V_2, G) + N_{\nu,\mu}^{EXP}(V_2, V_1, G)}{N_{tot}^{EXP}(V_1, V_2, G) + N_{tot}^{EXP}(V_2, V_1, G)} \quad \text{for geometries } B \text{ and } C \quad (8)$$

of the distributions measured in geometry $B$ and $C$ and in the equivalent geometries $A_1$ and $A_2$ was used for comparison with the simulations. In eqs. (7) and (8) $N_{\mu,\nu}^{EXP}(V_1, V_2, G)$ ($N_{\nu,\mu}^{EXP}(V_2, V_1, G)$) denote the number of pairs of ionisation clusters of size $\nu$ measured in $V_1$ and $\mu$ measured in $V_2$ ($\nu$ measured in $V_2$ and $\mu$ measured in $V_1$), and $N_{tot}^{EXP}(V_1, V_2, G) = N_{tot}^{EXP}(V_2, V_1, G)$ is the total number of ionisation cluster pairs measured in geometry $G$.

### 2.3 $\chi^2$ independence test and correlation coefficients

For the examination of the correlation of simulated ICSDs in the two sites, $V_1$ and $V_2$, a $\chi^2$ independence test was applied to the simulated bivariate distributions $P_{\mu,\nu}(V_1,V_2,G)$ using the product of the marginal distributions $P_\mu(V_1,G)$ and $P_\nu(V_2,G)$ as expected relative frequencies:

$$\chi^2(V_1, V_2, G) = \sum_{\mu=0}^{\mu_{max}} \sum_{\nu=0}^{\nu_{max}} \frac{\left(P_{\mu,\nu}(V_1, V_2, G) - P_\mu(V_1, G) \cdot P_\nu(V_2, G)\right)^2}{P_\mu(V_1, G) \cdot P_\nu(V_2, G)} \quad (9)$$

For estimation of the degree of correlation of the simulated ICSDs the empirical correlation coefficient $R(V_1,V_2,G)$ was calculated for $N$ particle tracks with $n = 1, \ldots, N$:

$$R(V_1, V_2, G) = \frac{\sum_{n=1}^{N}(\mu_n - M_1(V_1, G)) \cdot (\nu_n - M_1(V_2, G))}{\sqrt{\sum_{n=1}^{N}(\mu_n - M_1(V_1, G))^2 \cdot \sum_{n=1}^{N}(\nu_n - M_1(V_2, G))^2}} \quad (10)$$

where $\mu_n$ ($\nu_n$) is the ionisation cluster size produced by the $n$-th particle track in $V_1$ ($V_2$) and $M_1(V_1,G)$ ($M_2(V_2,G)$) is the mean ionisation cluster size produced by all $N$ particle tracks in $V_1$ ($V_2$) in geometry $G$.

For the cumulative probabilities $F_{k,k}(V_1,V_2,G)$ (see eq. (4) and eq. (11) in (Hilgers and Rabus, 2020)) the corresponding empirical correlation coefficient $R_{k,k}(V_1,V_2,G)$ was calculated for $N$ particle tracks, $n = 1, \ldots, N$:

$$R_{k,k}(V_1, V_2, G) = \frac{\sum_{n=1}^{N}(P(\mu_n) - F_k(V_1, G)) \cdot (P(\nu_n) - F_k(V_2, G))}{\sqrt{\sum_{n=1}^{N}(P(\mu_n) - F_k(V_1, G))^2 \cdot \sum_{n=1}^{N}(P(\nu_n) - F_k(V_2, G))^2}} \quad (11)$$

$$\text{with:} \quad P(\mu_n) = \begin{cases} 1 & \text{if: } \mu_n \geq k \\ 0 & \text{if: } \mu_n < k \end{cases} \quad \text{and:} \quad P(\nu_n) = \begin{cases} 1 & \text{if: } \nu_n \geq k \\ 0 & \text{if: } \nu_n < k \end{cases} \quad (12)$$

where $\mu_n$ ($\nu_n$) is the ionisation cluster size produced by the $n$-th particle track in $V_1$ ($V_2$) and $F_k(V_1,G)$ ($F_k(V_2,G)$) is the cumulative probability for all $N$ particle tracks to produce an ionisation cluster of size $k$ or larger in $V_1$ ($V_2$) in geometry $G$.

## 3. Results and discussion

Due to uncertainties of (i) the beam parameters (size, profile and position) in measurements and simulations, of (ii) the determination of the spatial distribution of the extraction efficiency and of (iii) the cross sections and different electron transport models in the simulations, the determination of the geometrical parameters $D_{eff}$, $H_{eff}$ and $D_{V_1V_2}$ as well as the ($\rho\lambda_{ion}$)-ratios are subject to uncertainties. These uncertainties are discussed in detail in (Hilgers et al., 2021) and can be transferred to the present analysis. Therefore, the uncertainty of the geometrical parameters $D_{eff}$, $H_{eff}$ and $D_{V_1V_2}$ as well as the uncertainty of the ($\rho\lambda_{ion}$)-ratios is estimated to ±15%.

### 3.1. Simulations using the spatial distribution of the extraction efficiency

#### 3.1.1. Simulations for $C_3H_8$ gas with PTra

Fig. 3 shows the simulated (left) and measured (right) probability distribution $P_{\nu,\mu}(V_1,V_2,A_1)$ for the simultaneous production of $\nu$ and $\mu$ ionisations in the corresponding sites $V_1$ and $V_2$ for 1.2 mbar $C_3H_8$. The alpha particles pass site $V_1$ centrally (geometry $A_1$). The measured distribution agrees well with the simulated, only minor deviations are found for larger ionisation clusters with relative frequencies of occurrence below $10^{-5}$. These deviations are due to a background of secondary ions (Hilgers and Rabus, 2019) leading to tails of additional large ionisation clusters.

The effect of this background becomes more visible in Fig. 3 for the simulated (left) and measured (right) probability distribution $P_{\nu,\mu}(V_1,V_2,B)$ with the primary ions passing the centre between $V_1$ and $V_2$ (geometry $B$). The shape of the measured distribution agrees well with the simulated for small cluster sizes. However, the tails of additional ionisation clusters at large cluster sizes are clearly visible, according to the finding in (Hilgers and Rabus, 2019) that ICSDs created by primaries passing by the site are stronger affected by the background of secondary ions than ICSDs created by primaries passing the site close to or at the centre.

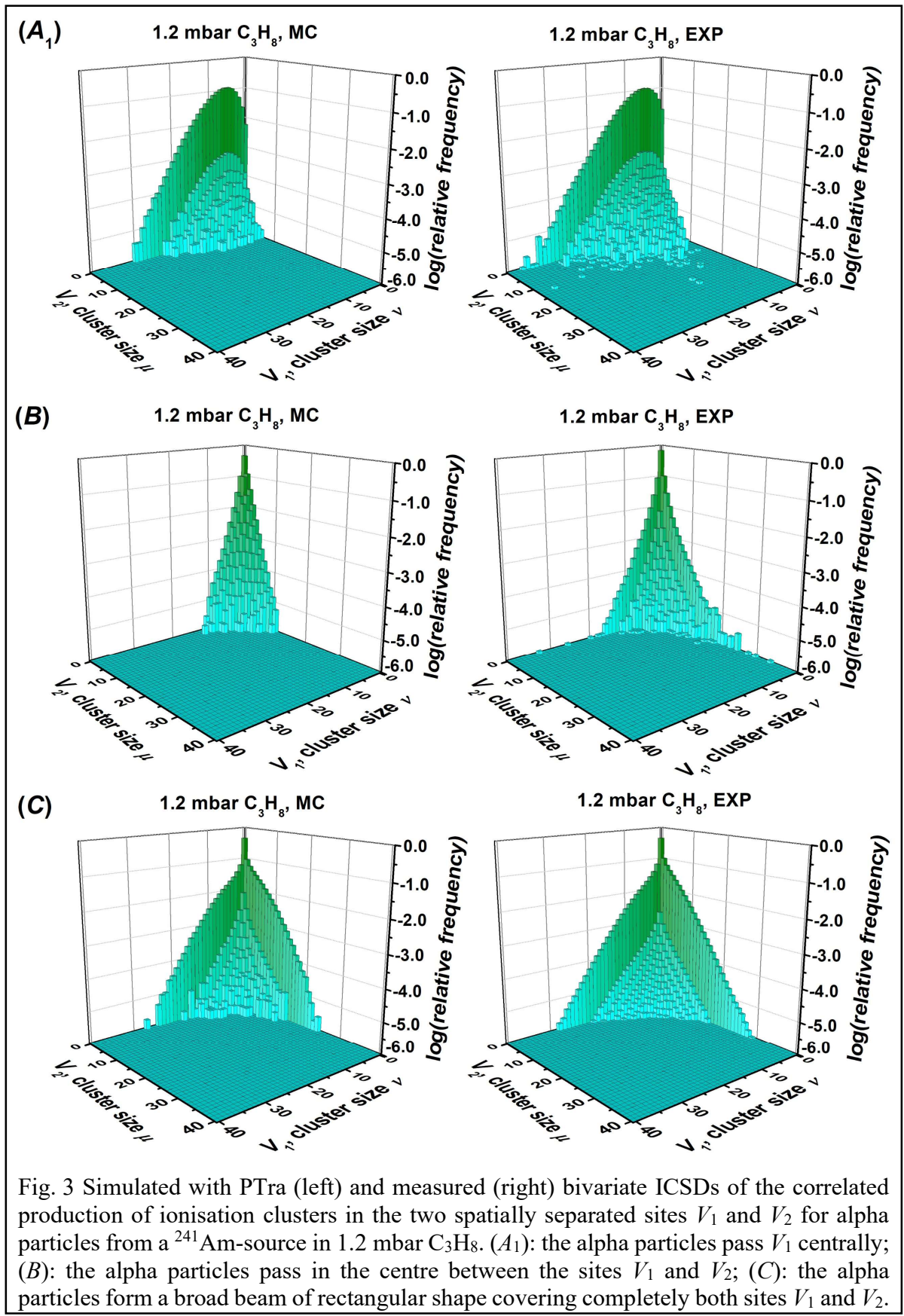

Fig. 3 Simulated with PTra (left) and measured (right) bivariate ICSDs of the correlated production of ionisation clusters in the two spatially separated sites $V_1$ and $V_2$ for alpha particles from a $^{241}$Am-source in 1.2 mbar $C_3H_8$. ($A_1$): the alpha particles pass $V_1$ centrally; (B): the alpha particles pass in the centre between the sites $V_1$ and $V_2$; (C): the alpha particles form a broad beam of rectangular shape covering completely both sites $V_1$ and $V_2$.

A good agreement can be seen between simulated and measured probability distributions $P_{\nu,\mu}(V_1,V_2,C)$ when the primary ion beam extends over an area covering completely both sites $V_1$ and $V_2$ (geometry $C$). In this case, only minor deviations are observed between simulation and measurement.

In total, the bivariate distributions of the simultaneous production of ionisation clusters for alpha particles from a $^{241}$Am-source in 1.2 mbar $C_3H_8$ simulated with PTra are found to agree very well with the corresponding measured distributions for the irradiation geometries investigated. The deviations between simulation and measurement can be attributed to a background of secondary ionisations, which results in tails of additional ionisation clusters at large cluster sizes.

### 3.1.2. Simulations in liquid H$_2$O with Geant4-DNA

Fig. 4 shows a comparison with measured data and the results of simulations with Geant4-DNA carried out using the spatial distributions of the extraction efficiency scaled in terms of liquid H$_2$O for the three target gases (a) 1.2 mbar H$_2$O, (b) 1.2 mbar $C_3H_8$ and (c) 1.2 mbar $C_4H_8O$ for alpha particles passing $V_1$ centrally (geometry $A_1$). On the whole, a good agreement between measurements and simulations is found for this geometry. For 1.2 mbar $C_3H_8$ and for 1.2 mbar $C_4H_8O$ minor deviations are found for large cluster sizes $\nu$ and $\mu$ showing smaller frequencies in the simulation than in the measurement. These deviations can again be attributed to a background of secondary ions (Hilgers and Rabus, 2019).

Fig. 5 shows the corresponding comparison of simulation results and experiments for alpha particles passing in the centre between the sites $V_1$ and $V_2$ (geometry $B$). At first glance, measurements and simulations seem to agree quite well. However, closer inspection reveals substantial differences in the shape of the ICSDs at small cluster sizes, which are most pronounced for $C_4H_8O$ and are least pronounced for H$_2$O. In the simulations, the peak at $\mu = \nu = 0$ is less pronounced than in the measurements, i.e., the relative frequency of cluster sizes $\mu, \nu \cong 0$ is lower than that found in the measurements. However, at small cluster sizes around $\mu, \nu \cong 1 - 3$ the relative frequencies are larger in the simulations compared to those found in the measurements.

Table 1 shows the values of $\chi^2 / n$ and $\kappa$ obtained for the comparison between the measured and simulated ICSDs for the three target gases using Geant4-DNA and PTra (only for $C_3H_8$) for the three geometries. For the data denoted with "(G4-DNA)" the ICSDs were obtained by superposition of the tracks simulated in liquid H$_2$O with the spatial distributions of the extraction efficiency that were simulated for the respective gas and then scaled to liquid H$_2$O. Comparison of the values of $\chi^2 / n$ and $\kappa$ obtained for the three different geometries show significantly larger values of both $\chi^2 / n$ and $\kappa$ for geometry $B$ than for the two other geometries. This can be attributed to the procedure (Grosswendt, 2006) applied to scale the dimensions of the spatial distribution of the extraction efficiency. This procedure is based, besides the density of the materials involved, on the ratio of the mean free paths for ionisation of the alpha particles in the respective materials and is, therefore, only valid for ionisations due to the primary alpha particles and not for the secondary electrons in the penumbra of the primary particle track. However, in geometry $B$ the ionisations in $V_1$ and $V_2$ are practically exclusively due to ionisations of secondary electrons, since the primary alpha particles pass through the centre point between the centres of $V_1$ and $V_2$ (see Fig. 2). Therefore, the scaling procedure is not valid in geometry $B$, which is reflected in the large values of $\chi^2 / n$ and $\kappa$, in contrast to geometries $A_1$ ($A_2$) and $C$, where the ICSDs contain substantial contributions of ionisations due to the primary alpha particles.

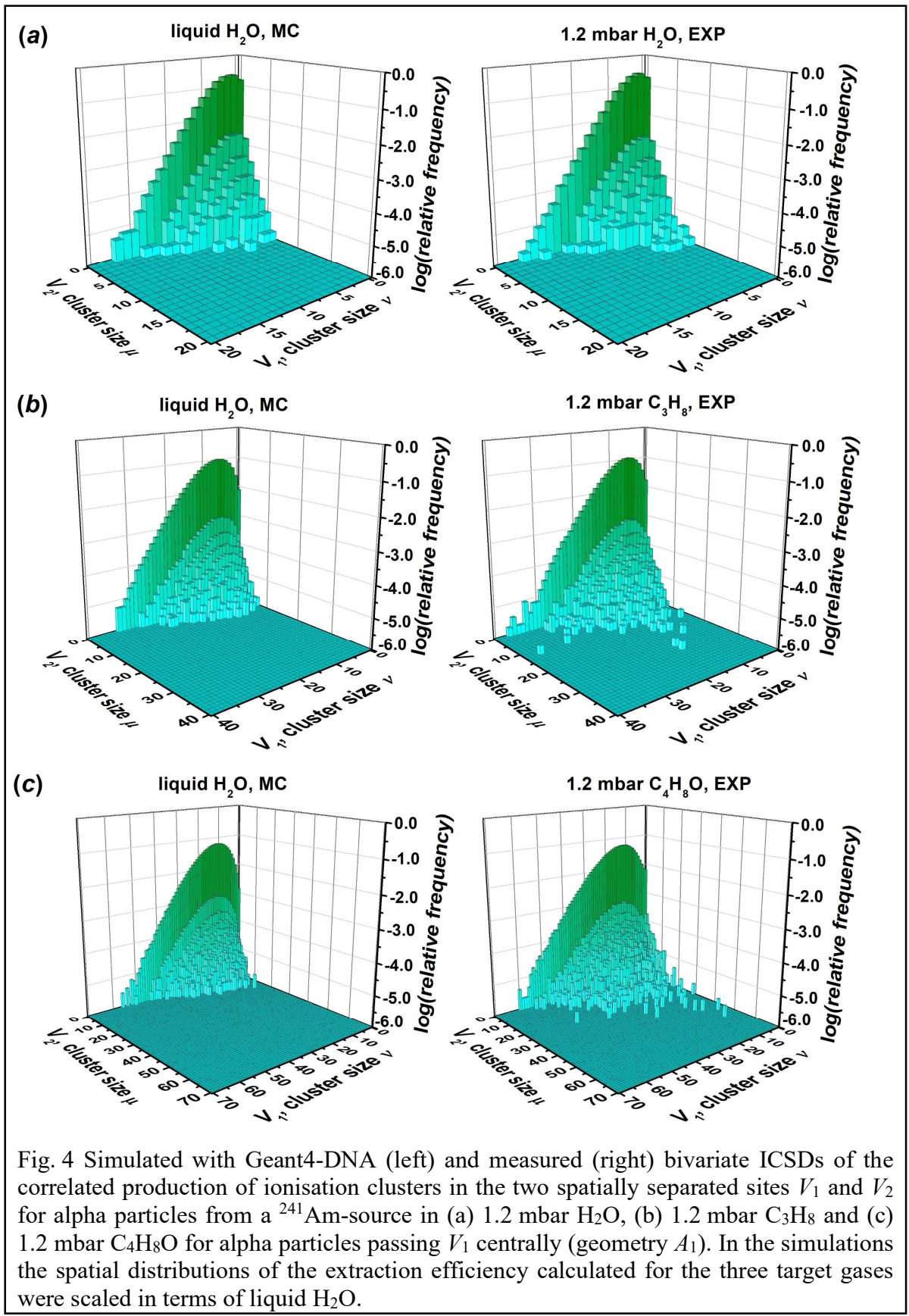

Fig. 4 Simulated with Geant4-DNA (left) and measured (right) bivariate ICSDs of the correlated production of ionisation clusters in the two spatially separated sites $V_1$ and $V_2$ for alpha particles from a $^{241}$Am-source in (a) 1.2 mbar $H_2O$, (b) 1.2 mbar $C_3H_8$ and (c) 1.2 mbar $C_4H_8O$ for alpha particles passing $V_1$ centrally (geometry $A_1$). In the simulations the spatial distributions of the extraction efficiency calculated for the three target gases were scaled in terms of liquid $H_2O$.

For geometries $A_1$ and $C$ no common trend for $\chi^2/n$ and $\kappa$ is observed. For geometry $A_1$ $\chi^2/n$ tends to smaller values, whereas $\kappa$ is clearly smaller for geometry $C$. With respect to the target gas a general trend for $\chi^2/n$ and $\kappa$ is also not observed. From PTra simulations for $C_3H_8$ both $\chi^2/n$ and $\kappa$ are significantly smaller than those obtained for simulations with Geant4-DNA (except for $\chi^2/n$ in geometry $A_1$), indicating a better agreement between measurement and simulation for the simulations using PTra. This is expected, since in PTra simulations are carried out in $C_3H_8$ gas.

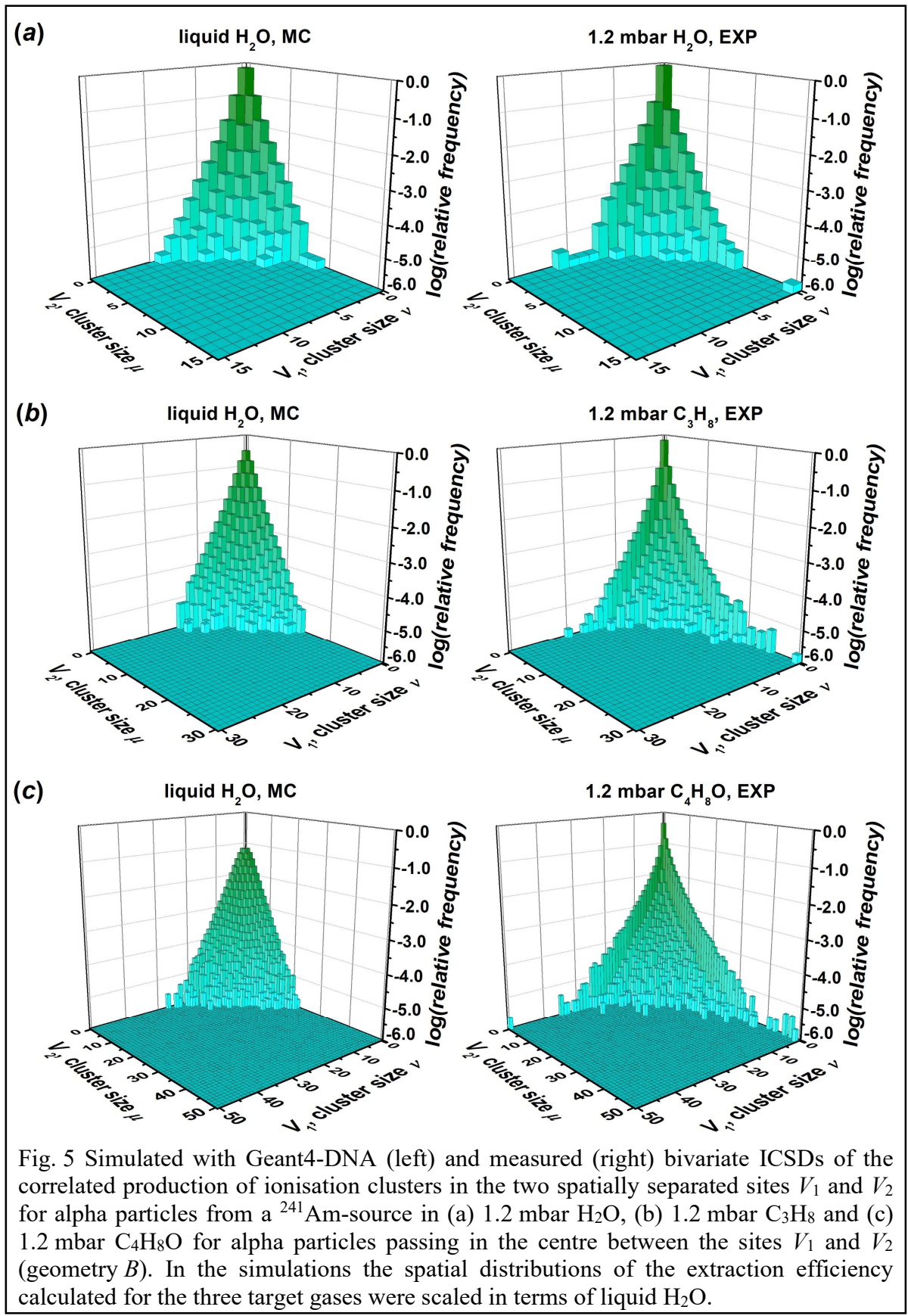

Fig. 5 Simulated with Geant4-DNA (left) and measured (right) bivariate ICSDs of the correlated production of ionisation clusters in the two spatially separated sites $V_1$ and $V_2$ for alpha particles from a $^{241}$Am-source in (a) 1.2 mbar $H_2O$, (b) 1.2 mbar $C_3H_8$ and (c) 1.2 mbar $C_4H_8O$ for alpha particles passing in the centre between the sites $V_1$ and $V_2$ (geometry $B$). In the simulations the spatial distributions of the extraction efficiency calculated for the three target gases were scaled in terms of liquid $H_2O$.

|  | $A_1$ | | B | | C | |
|---|---|---|---|---|---|---|
|  | $\chi^2/n$ | $\kappa$ | $\chi^2/n$ | $\kappa$ | $\chi^2/n$ | $\kappa$ |
| $H_2O$ (G4-DNA) | 116 | 0.064 | 1979 | 0.085 | 180 | 0.043 |
| $C_3H_8$ (G4-DNA) | 64.6 | 0.092 | 2654 | 0.305 | 80.5 | 0.067 |
| $C_3H_8$ (PTra) | 70.2 | 0.070 | 1909 | 0.188 | 33.3 | 0.013 |
| $C_4H_8O$ (G4-DNA) | 19.4 | 0.064 | 1615 | 0.434 | 47.4 | 0.036 |

Table 1. $\chi^2/n$ and $\kappa$ for the comparison between the measured and simulated ICSDs for the three target gases using Geant4-DNA (G4-DNA) and PTra (only for $C_3H_8$) for the three geometries $A_1$, $B$, and $C$. For the data denoted with "(G4-DNA)" the ICSDs were obtained by superposition of the tracks simulated in liquid $H_2O$ with the spatial distributions of the extraction efficiency that were simulated for the respective gas and then scaled to liquid $H_2O$.

The large numerical values of $\chi^2/n$ are due to the inclusion of only the statistical uncertainties of $P_{\mu,\nu}^{EXP}(G)$ and $P_{\mu,\nu}^{MC}(G)$ in eq. (5). The sum of the squares of the uncertainties builds the denominator in eq. (5), which is numerically much smaller than the numerator, thus increasing the numerical value of each summand of the sum in eq. (5).

### 3.2. Optimisation for cylindrical targets using the marginal distributions

According to the discussion in section 2.2., the parameters describing the cylindrical volume, $D_{eff}$ and $H_{eff}$, were optimised in the first step using the arithmetic mean of the two marginal distributions of primary alpha particles passing the target cylinders $V_1$ and $V_2$ centrally (geometry $A_1$ and $A_2$). Additionally, a proportional relation $H_{eff} = F \cdot D_{eff}$ was assumed with $F = H_{eff}/D_{eff}$ resulting from $D_{eff}$ and $H_{eff}$ obtained by integration of the extraction efficiency. After optimising $D_{eff}$ (and, with it, $H_{eff}$), a similar procedure was carried out to optimise $D_{V_1V_2}$ in the second step keeping the previously optimised $D_{eff}$ and $H_{eff}$. Here the simulations were performed for a single cylinder with the alpha particles passing "below" the bottom of the cylinder (PTra) or for a pair of congruent cylinders with irradiation geometry $B$ in Fig. 2 (Geant4-DNA), and the experimental reference data was the arithmetic mean of the marginal distributions in the two targets for the case that the alpha particle beam passed centrally between the two targets.

### 3.2.1. Simulations for $C_3H_8$ gas with PTra

Comparison of ICSDs measured in 1.2 mbar $C_3H_8$ and the corresponding ICSDs simulated with PTra using an ideal cylinder of dimensions $D_{eff}$ and $H_{eff}$ as target volume can be seen in Fig. 6, which shows contour plots of the corresponding $\kappa(D_{eff},H_{eff})$ (left) and $\chi^2(D_{eff},H_{eff})/n$ (right). The red circle with the white centre marks $D_{eff}$ and $H_{eff}$ obtained by integrating the spatial distribution of the extraction efficiency. Along the red line, the ratio $F = H_{eff}/D_{eff}$ is identical to that resulting from $D_{eff}$ and $H_{eff}$ obtained by integration of the spatial distribution of the extraction efficiency. The dashed red lines mark the interval band for $D_{eff}$ and $H_{eff}$ due to the uncertainty of $F$. The dotted vertical lines mark that $D_{eff}$ for which the deviation between measurement and simulation is minimal, if the ratio $F = H_{eff}/D_{eff}$ is applied.

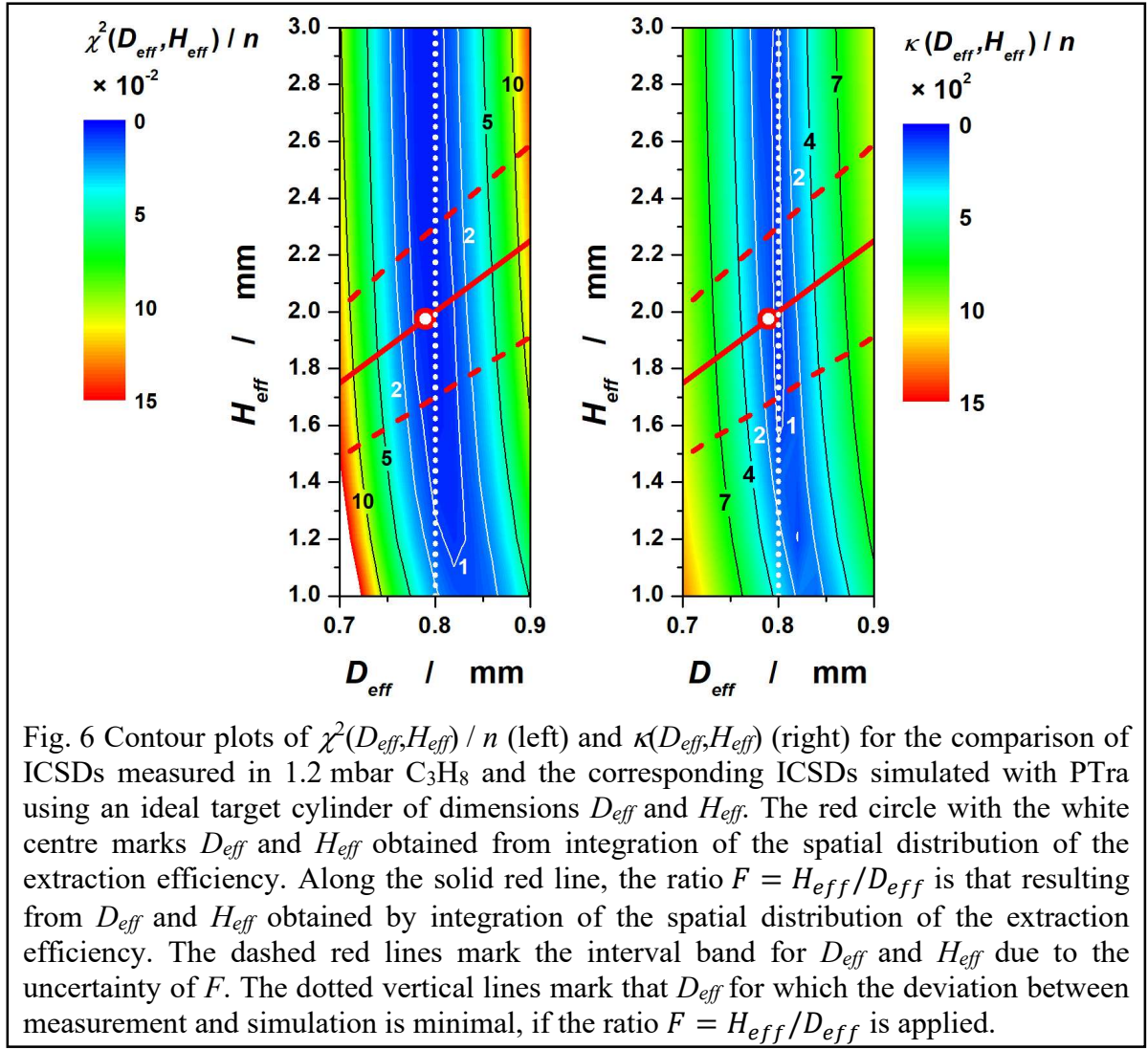

Fig. 6 Contour plots of $\chi^2(D_{eff},H_{eff}) / n$ (left) and $\kappa(D_{eff},H_{eff})$ (right) for the comparison of ICSDs measured in 1.2 mbar $C_3H_8$ and the corresponding ICSDs simulated with PTra using an ideal target cylinder of dimensions $D_{eff}$ and $H_{eff}$. The red circle with the white centre marks $D_{eff}$ and $H_{eff}$ obtained from integration of the spatial distribution of the extraction efficiency. Along the solid red line, the ratio $F = H_{eff}/D_{eff}$ is that resulting from $D_{eff}$ and $H_{eff}$ obtained by integration of the spatial distribution of the extraction efficiency. The dashed red lines mark the interval band for $D_{eff}$ and $H_{eff}$ due to the uncertainty of $F$. The dotted vertical lines mark that $D_{eff}$ for which the deviation between measurement and simulation is minimal, if the ratio $F = H_{eff}/D_{eff}$ is applied.

Similar to the findings in (Hilgers et al., 2021), the dependence of $\kappa$ and $\chi^2 / n$ on $H_{eff}$ is much less pronounced than that on $D_{eff}$, i.e. the simulated ICSDs vary only slightly with varying $H_{eff}$, whereas small variations of $D_{eff}$ lead to significant changes in $\kappa$ and $\chi^2 / n$. For small $H_{eff}$ there is also a stronger dependence on $H_{eff}$, such that a variation in $D_{eff}$ can be compensated to some extent by an appropriate change in $H_{eff}$. Therefore, a large variety of values of $D_{eff}$ and $H_{eff}$ leads to small $\chi^2 / n$ and $\kappa$, making a unique determination of $D_{eff}$ and $H_{eff}$ impossible. However, the integration of the spatial distribution of the extraction efficiency yields additional information on $D_{eff}$ and $H_{eff}$, in particular the ratio $F = H_{eff}/D_{eff}$. Applying this ratio in the comparison of measured and simulated ICSDs leads to unique results for $D_{eff}$ and $H_{eff}$ which are consistent with the shape of $\chi^2 / n$ and $\kappa$. Therefore, the application of the ratio $F = H_{eff}/D_{eff}$ obtained by integration of the spatial distribution of the extraction efficiency seems justified. The point of $D_{eff}$ and $H_{eff}$ obtained by integration of the spatial distribution of the extraction efficiency lies almost at the bottom of the valleys of $\kappa$ and $\chi^2 / n$.

Assuming the ratio of $F = H_{eff}/D_{eff} = 2.5$ as obtained by integration of the spatial distribution of the extraction efficiency (Hilgers and Rabus, 2019), at first the values of $D_{eff}$ and $H_{eff}$ are varied according to this ratio as indicated by the red lines in the plots of Fig. 6, i.e. $H_{eff} = 2.5 \cdot D_{eff}$, and after that the distance of the alpha particles passing below the bottom of the cylinder (which is related to $D_{V_1V_2}$). The result of these two optimisations is shown in Fig. 7. The left plot shows the comparison between the measured and simulated ICSDs calculated using those values for the parameters $D_{eff}$ (and $H_{eff} = 2.5 \cdot D_{eff}$) and $D_{V_1V_2}$ leading to minima of $\kappa(D_{eff}, H_{eff} = 2.5 \cdot D_{eff})$ and of $\kappa(D_{V_1V_2})$. In the measurements, the mean of the two measured distributions is shown. Very good agreement between measurement and simulation is found for both geometry $A$, where the alpha particles pass the target centrally, and for geometry $B$, where the alpha particles pass the target at a distance from the target centre, but on the central axis of the target cylinder.

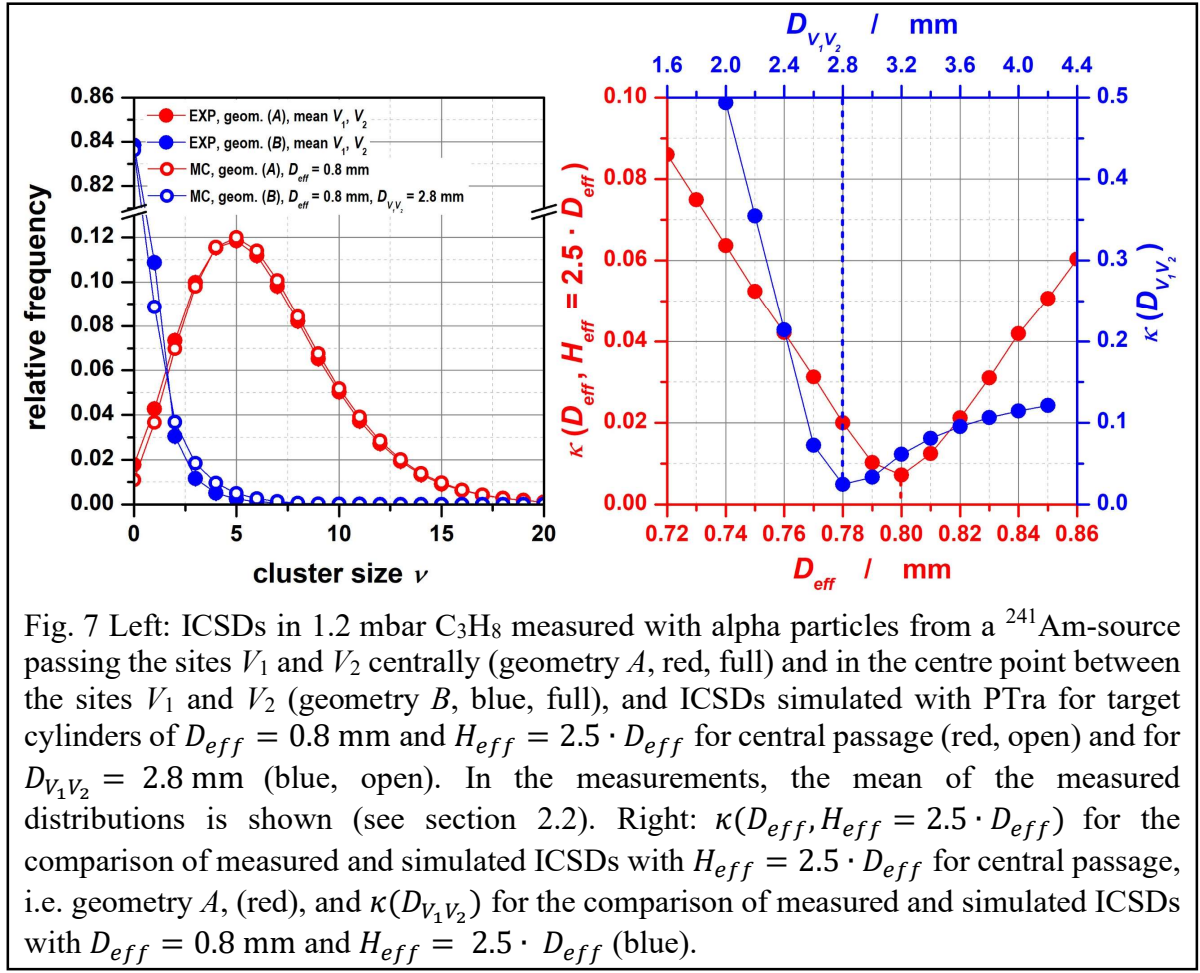

Fig. 7 Left: ICSDs in 1.2 mbar $C_3H_8$ measured with alpha particles from a $^{241}$Am-source passing the sites $V_1$ and $V_2$ centrally (geometry $A$, red, full) and in the centre point between the sites $V_1$ and $V_2$ (geometry $B$, blue, full), and ICSDs simulated with PTra for target cylinders of $D_{eff} = 0.8$ mm and $H_{eff} = 2.5 \cdot D_{eff}$ for central passage (red, open) and for $D_{V_1V_2} = 2.8$ mm (blue, open). In the measurements, the mean of the measured distributions is shown (see section 2.2). Right: $\kappa(D_{eff}, H_{eff} = 2.5 \cdot D_{eff})$ for the comparison of measured and simulated ICSDs with $H_{eff} = 2.5 \cdot D_{eff}$ for central passage, i.e. geometry $A$, (red), and $\kappa(D_{V_1V_2})$ for the comparison of measured and simulated ICSDs with $D_{eff} = 0.8$ mm and $H_{eff} = 2.5 \cdot D_{eff}$ (blue).

In Fig. 7 only $\kappa(D_{eff}, H_{eff} = 2.5 \cdot D_{eff})$ and of $\kappa(D_{V_1V_2})$ for $D_{eff} = 0.8$ mm are shown. The corresponding minima of $\chi^2(D_{eff}, H_{eff} = 2.5 \cdot D_{eff})/n$ and $\chi^2(D_{V_1V_2})/n$ are found at the same positions.

The effective diameter $D_{eff} = 0.8$ mm obtained by minimisation of $\kappa(D_{eff}, H_{eff} = 2.5 \cdot D_{eff})$ differs from $D_{eff} = 0.79$ mm obtained by integration of the extraction efficiency by less than 1.5%. Thus, for 1.2 mbar $C_3H_8$ the measuring geometry can be represented by two cylinders stacked one above the other along their central axis with $D_{eff} = 0.8$ mm and $H_{eff} = 2.5 \cdot D_{eff} = 2.0$ mm. The distance between both cylinder centres is $D_{V_1V_2} = 2.8$ mm, i.e. the distance between the top of $V_1$ and the bottom of $V_2$ is 0.8 mm.

### 3.2.2. Simulations in liquid $H_2O$ with Geant4-DNA

Integration of the spatial distribution of the extraction efficiency results in $F = H_{eff}/D_{eff} = 2.6$ for 1.2 mbar $H_2O$, $F = 2.5$ for 1.2 mbar $C_3H_8$, and $F = 3.0$ for 1.2 mbar $C_4H_8O$, averaged over the two sites $V_1$ and $V_2$ (Hilgers and Rabus, 2020). For better comparison, the effective diameters and distances are all scaled to liquid $H_2O$ (index *lw*) at unit density. Applying the scaling factors from (Hilgers et al., 2021) leads, therefore, to $D_{eff,lw} = 1.0$ nm for 1.2 mbar $H_2O$, $D_{eff,lw} = 2.5$ nm for 1.2 mbar $C_3H_8$, and $D_{eff,lw} = 4.0$ nm for 1.2 mbar $C_4H_8O$.

For the determination of $D_{eff,lw}$, $H_{eff,lw}$ and $D_{V_1V_2,lw}$ for $H_2O$, $C_3H_8$ and $C_4H_8O$, the same procedure was carried out as before for the determination of $D_{eff}$, $H_{eff}$ and $D_{V_1V_2}$ for $C_3H_8$: first $D_{eff,lw}$ was varied in the simulations assuming the ratio $F = H_{eff,lw}/D_{eff,lw}$ for the respective target gas obtained by integration of the spatial distribution of the extraction efficiency and in the second step $D_{V_1V_2,lw}$ was varied for that $D_{eff,lw}$ which minimised $\kappa(D_{eff,lw}, H_{eff,lw} = F \cdot D_{eff,lw})$.

The results of this determination are shown in Fig. 8 for the three target gases. Measurements and simulations agree very well. For $H_2O$ the effective diameters $D_{eff,lw} = 1.0$ nm obtained by scaling of the integrated spatial distribution of the extraction efficiency and $D_{eff,lw} = 0.925$ nm determined by minimisation of $\kappa(D_{eff,lw}, H_{eff,lw} = 2.6 \cdot D_{eff,lw})$ differ by ≈8%, and the distance between the cylinder centres is $D_{V_1V_2,lw} = 4.0$ nm, resulting in a geometry of two cylinders of liquid $H_2O$ of 0.925 nm in diameter and 2.4 nm in height (assuming $F = H_{eff,lw}/D_{eff,lw} = 2.6$), stacked one above the other along the central axis of the two cylinders with the centres of the two cylinders being 4.0 nm apart, i.e. with a distance between the top of $V_1$ and the bottom of $V_2$ of 1.6 nm.

For $C_3H_8$ the effective diameters $D_{eff,lw} = 2.5$ nm obtained by scaling of the integrated spatial distribution of the extraction efficiency, and $D_{eff,lw} = 2.45$ nm determined by minimisation of $\kappa(D_{eff,lw}, H_{eff,lw} = F \cdot D_{eff,lw})$ differ by ≈2%, the distance between the cylinder centres is $D_{V_1V_2,lw} = 10.5$ nm. The resulting geometry consists of two cylinders of liquid $H_2O$ of 2.45 nm in diameter and 6.1 nm in height (assuming $F = H_{eff,lw}/D_{eff,lw} = 2.5$), stacked one above the other along the central axis of the two cylinders with the centres of the two cylinders being 10.5 nm apart, i.e. with a distance between the top of $V_1$ and the bottom of $V_2$ of 4.4 nm.

For $C_4H_8O$ the effective diameters $D_{eff,lw} = 4.0$ nm obtained by scaling of the integrated spatial distribution of the extraction efficiency and $D_{eff,lw} = 4.5$ nm determined by minimisation of $\kappa(D_{eff,lw}, H_{eff,lw} = F \cdot D_{eff,lw})$ differ by less than ≈13%, and for the distance between the cylinder centre and the beam axis $D_{V_1V_2,lw} = 19$ nm is found, leading to a geometry of two cylinders of liquid $H_2O$ of 4.5 nm in diameter and 13.5 nm in height (assuming $F = H_{eff,lw}/D_{eff,lw} = 3.0$), stacked one above the other along the central axis of the two cylinders with the centres of the two cylinders being 19.0 nm apart, i.e. with a distance between the top of $V_1$ and the bottom of $V_2$ of 5.5 nm.

Table 2 summarizes the optimised the parameters $D_{eff,lw}$, $H_{eff,lw}$ and $D_{V_1V_2,lw}$ for the three target gases 1.2 mbar $H_2O$, 1.2 mbar $C_3H_8$ and 1.2 mbar $C_4H_8O$.

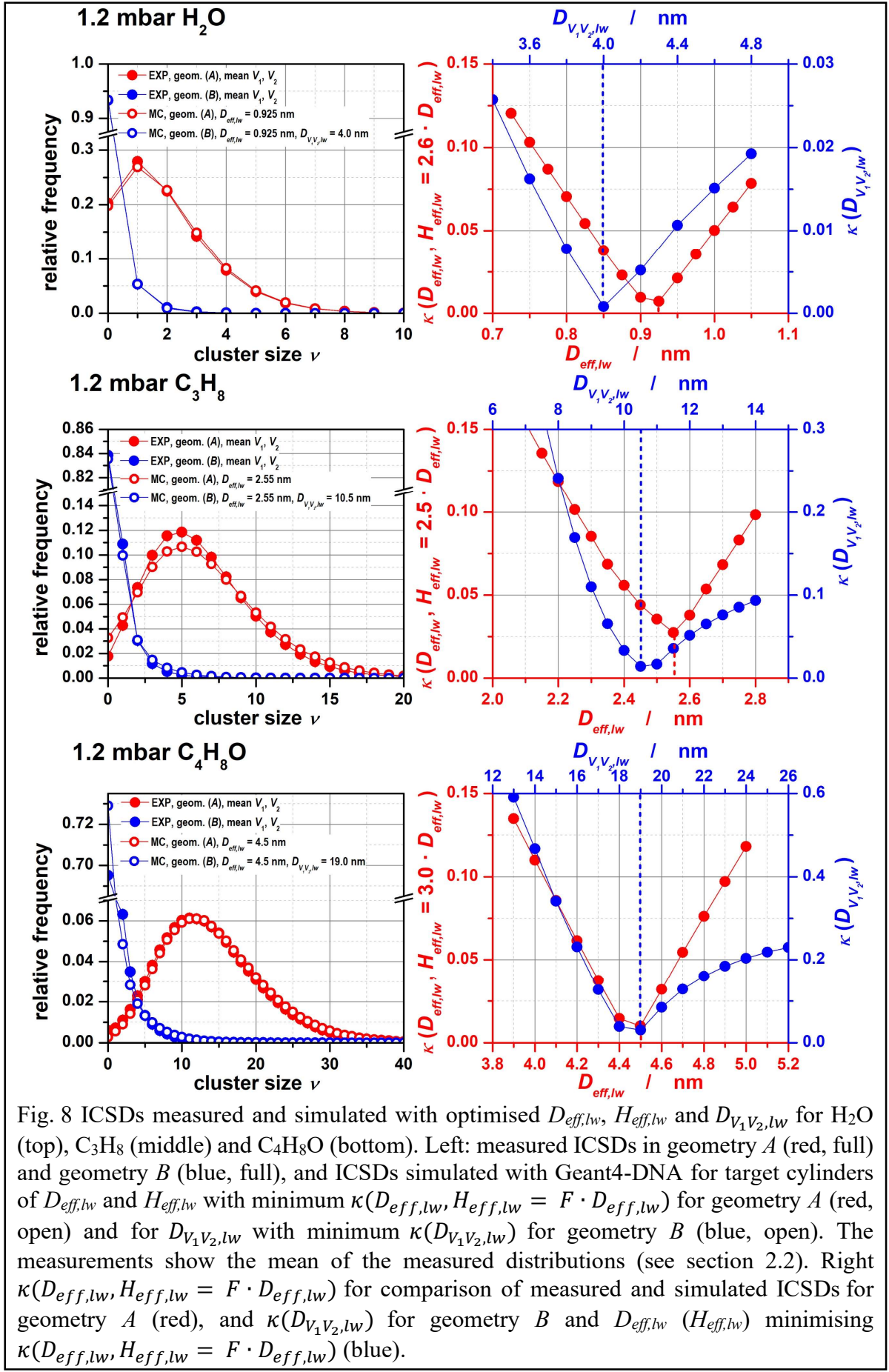

Fig. 8 ICSDs measured and simulated with optimised $D_{eff,lw}$, $H_{eff,lw}$ and $D_{V_1V_2,lw}$ for $H_2O$ (top), $C_3H_8$ (middle) and $C_4H_8O$ (bottom). Left: measured ICSDs in geometry $A$ (red, full) and geometry $B$ (blue, full), and ICSDs simulated with Geant4-DNA for target cylinders of $D_{eff,lw}$ and $H_{eff,lw}$ with minimum $\kappa(D_{eff,lw}, H_{eff,lw} = F \cdot D_{eff,lw})$ for geometry $A$ (red, open) and for $D_{V_1V_2,lw}$ with minimum $\kappa(D_{V_1V_2,lw})$ for geometry $B$ (blue, open). The measurements show the mean of the measured distributions (see section 2.2). Right $\kappa(D_{eff,lw}, H_{eff,lw} = F \cdot D_{eff,lw})$ for comparison of measured and simulated ICSDs for geometry $A$ (red), and $\kappa(D_{V_1V_2,lw})$ for geometry $B$ and $D_{eff,lw}$ ($H_{eff,lw}$) minimising $\kappa(D_{eff,lw}, H_{eff,lw} = F \cdot D_{eff,lw})$ (blue).

As in Fig. 7, in Fig. 8 only $\kappa(D_{eff,lw}, H_{eff,lw} = F \cdot D_{eff,lw})$ and of $\kappa(D_{V_1V_2,lw})$ for $D_{eff,lw}$

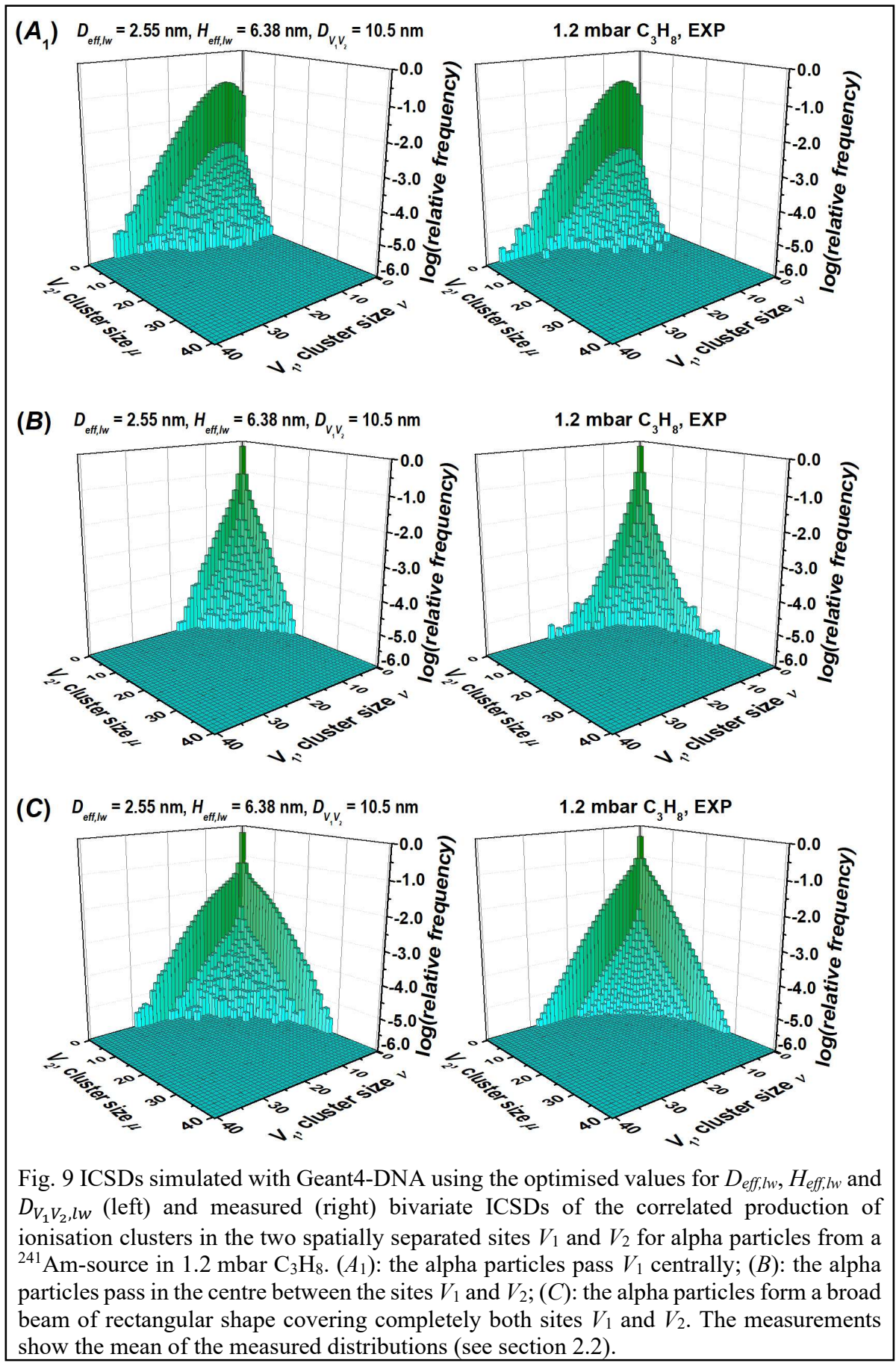

Fig. 9 ICSDs simulated with Geant4-DNA using the optimised values for $D_{eff,lw}$, $H_{eff,lw}$ and $D_{V_1V_2,lw}$ (left) and measured (right) bivariate ICSDs of the correlated production of ionisation clusters in the two spatially separated sites $V_1$ and $V_2$ for alpha particles from a $^{241}$Am-source in 1.2 mbar $C_3H_8$. ($A_1$): the alpha particles pass $V_1$ centrally; (B): the alpha particles pass in the centre between the sites $V_1$ and $V_2$; (C): the alpha particles form a broad beam of rectangular shape covering completely both sites $V_1$ and $V_2$. The measurements show the mean of the measured distributions (see section 2.2).

minimising $\kappa(D_{eff,lw}, H_{eff,lw} = F \cdot D_{eff,lw})$ are shown. The corresponding minima of $\chi^2(D_{eff,lw}, H_{eff,lw} = F \cdot D_{eff,lw})/n$ and $\chi^2(D_{V_1V_2,lw})/n$ are found at the same positions except for $C_3H_8$ with minima of $\kappa(D_{eff,lw}, H_{eff,lw} = F \cdot D_{eff,lw})$ at $D_{eff,lw} = 2.55$ nm and $\chi^2(D_{eff,lw}, H_{eff,lw} = F \cdot D_{eff,lw})/n$ at $D_{eff,lw} = 2.5$ nm.

|        | $D_{eff,lw}$ / nm | $H_{eff,lw}$ / nm | $D_{V_1V_2,lw}$ / nm |
|--------|-------------------|-------------------|----------------------|
| $H_2O$ | 0.925 | 2.41 | 4.0 |
| $C_3H_8$ | 2.55 | 6.38 | 10.5 |
| $C_4H_8O$ | 4.5 | 13.5 | 19.0 |

Table 2. Optimised parameters $D_{eff,lw}$, $H_{eff,lw}$ and $D_{V_1V_2,lw}$ of the cylindrical target volumes $V_1$ and $V_2$ for the three target gases 1.2 mbar $H_2O$, 1.2 mbar $C_3H_8$ and 1.2 mbar $C_4H_8O$. The estimated relative uncertainties are ±15%.

Fig. 9 shows simulations carried out with Geant4-DNA using the optimised values for $D_{eff,lw}$, $H_{eff,lw}$ and $D_{V_1V_2,lw}$ for alpha particles from a $^{241}$Am-source in 1.2 mbar $C_3H_8$ for the three geometries $A_1$, $B$ and $C$ together with the corresponding measurements. The measurements represent the mean of the measured distributions according to the discussion in section 2.2. Overall, a good agreement between measurements and simulations is found for all geometries. For geometry $B$ minor deviations are found for large cluster sizes $\nu$ and $\mu$ showing smaller frequencies in the simulation than in the measurement. These deviations can be attributed to a background of secondary ions (Hilgers and Rabus, 2019).

Table 3 shows the values of $\chi^2/n$ and $\kappa$ obtained for the comparison between the measured and simulated ICSDs for the three target gases with Geant4-DNA using the optimised parameters $D_{eff,lw}$, $H_{eff,lw}$ and $D_{V_1V_2,lw}$ of the cylindrical target volumes $V_1$ and $V_2$ for the three geometries $A$, $B$, and $C$. Comparison with Table 1, which shows data obtained for the comparison with simulations using the spatial distributions of the extraction efficiency scaled to liquid $H_2O$, shows for all target gases a clearly better agreement between measurement and simulation for geometry $A$ and $A_1$, respectively, and a substantially better agreement for geometry $B$. For geometry $C$, however, no clear trend is visible: for the simulation using the optimised parameters for cylindrical target volumes the agreement with the measurement is slightly better for $H_2O$, whereas simulations using the spatial distributions of the extraction efficiency scaled to liquid $H_2O$ agree slightly better with the measurement for $C_4H_8O$ and significantly better for $C_3H_8$.

|        | A | | B | | C | |
|--------|-----|-----|-----|-----|-----|-----|
|        | $\chi^2/n$ | $\kappa$ | $\chi^2/n$ | $\kappa$ | $\chi^2/n$ | $\kappa$ |
| $H_2O$ | 14.8 | 0.009 | 3.59 | 0.001 | 138 | 0.015 |
| $C_3H_8$ | 32.4 | 0.032 | 38.7 | 0.025 | 302 | 0.127 |
| $C_4H_8O$ | 2.16 | 0.011 | 31.6 | 0.051 | 63.8 | 0.099 |

Table 3. $\chi^2/n$ and $\kappa$ for the comparison between the measured and simulated ICSDs for the three target gases with Geant4-DNA using the optimised parameters $D_{eff,lw}$, $H_{eff,lw}$ and $D_{V_1V_2,lw}$ of the cylindrical target volumes $V_1$ and $V_2$ for the three geometries $A$, $B$, and $C$. The ICSDs were obtained by superposition of the tracks simulated in liquid $H_2O$ with the respective target cylinders of liquid $H_2O$.

### 3.3. Optimisation for cylindrical targets using bivariate distributions

As explained in section 2.2., the parameters $D_{eff,lw}$, $H_{eff,lw}$ and $D_{V_1V_2,lw}$ describing the geometry were also determined by an alternative approach in which they were optimised simultaneously. This approach was applied only to simulations using Geant4-DNA, since the present version of PTra does not allow the simulation of two stacked target cylinders. As before, a proportional relation $H_{eff,lw} = F \cdot D_{eff,lw}$ was assumed with the ratio $F = H_{eff,lw}/D_{eff,lw}$ being identical to that resulting from $D_{eff,lw}$ and $H_{eff,lw}$ obtained by integration of the spatial distribution of the extraction efficiency.

Fig. 10 shows the contour plots of $\chi^2(D_{eff,lw}, H_{eff,lw} = F \cdot D_{eff,lw}, D_{V_1V_2,lw})/n$ (left) and $\kappa(D_{eff,lw}, H_{eff,lw} = F \cdot D_{eff,lw}, D_{V_1V_2,lw})$ (right) for the comparison of ICSDs measured and simulated with Geant4-DNA for $H_2O$ (top), $C_3H_8$ (middle) and $C_4H_8O$ (bottom), assuming a proportional relation $H_{eff,lw} = F \cdot D_{eff,lw}$. The data show the arithmetic mean of $\chi^2/n$ and $\kappa$ obtained for the three geometries $A$, $B$ and $C$. The dotted lines indicate the values for parameters $D_{eff,lw}$ and $D_{V_1V_2,lw}$ obtained using the marginal distributions (section 3.2.2). For better readability, the index "$lw$" is omitted in the legend of the contour plots.

For $H_2O$, the position of the minima of $\chi^2/n$ and $\kappa$ in the contour plots agree well with the minima obtained previously for $D_{eff,lw}$ and $D_{V_1V_2,lw}$. For $C_3H_8$ and $C_4H_8O$, the position of the minima of $\chi^2/n$ in the contour plots is shifted by ≲5% towards smaller values of $D_{eff,lw}$ and $D_{V_1V_2,lw}$ compared to the minima obtained previously, whereas the minima of $\kappa$ are shifted by ≲3% towards larger values for $C_3H_8$ and towards smaller values for $C_4H_8O$.

The values found for $D_{eff,lw}$ and $D_{V_1V_2,lw}$ with the two different approaches agree within their estimated uncertainties. Therefore, both approaches can be regarded as equivalent procedures. However, it has to be pointed out that the procedure using the entire bivariate distribution is substantially more time consuming due to the significantly larger number of ICSDs, which have to be simulated.

### 3.4. Correlations between ionisation cluster size distributions of stacked target cylinders

Analogously to the investigation of correlations in measured bivariate ICSDs in (Hilgers and Rabus, 2020), for examination of the correlation of simulated ICSDs in the two sites, $V_1$ and $V_2$, a $\chi^2$ independence test was applied (see eq. (9)) to the simulated bivariate distributions $P_{\nu,\mu}(V_1,V_2,G)$ using the product of the marginal distributions as expected relative frequencies. To account for the differences of $V_1$ and $V_2$ in the simulations using the spatial distribution of the extraction efficiency scaled in terms of liquid $H_2O$, the bivariate frequency distributions obtained for the equivalent geometries $A_1$ and $A_2$ were averaged according to section 2.2.

Due to the large number of low frequency events in the bivariate distributions, a subset of $P_{\nu,\mu}(V_1,V_2,G)$ was used to construct the contingency table, where complementary cumulative frequencies were used in the last column $\nu_{max}$ and row $\mu_{max}$, respectively. The values of $\nu_{max}$ and $\mu_{max}$ were determined from the requirement that the fraction of expected frequency values in the contingency table that are smaller than 5 is below 5% and that the product of the degrees of freedom and of the sum of 'expected' event frequencies for $\nu \leq \nu_{max}$ and $\mu \leq \mu_{max}$ becomes maximal.

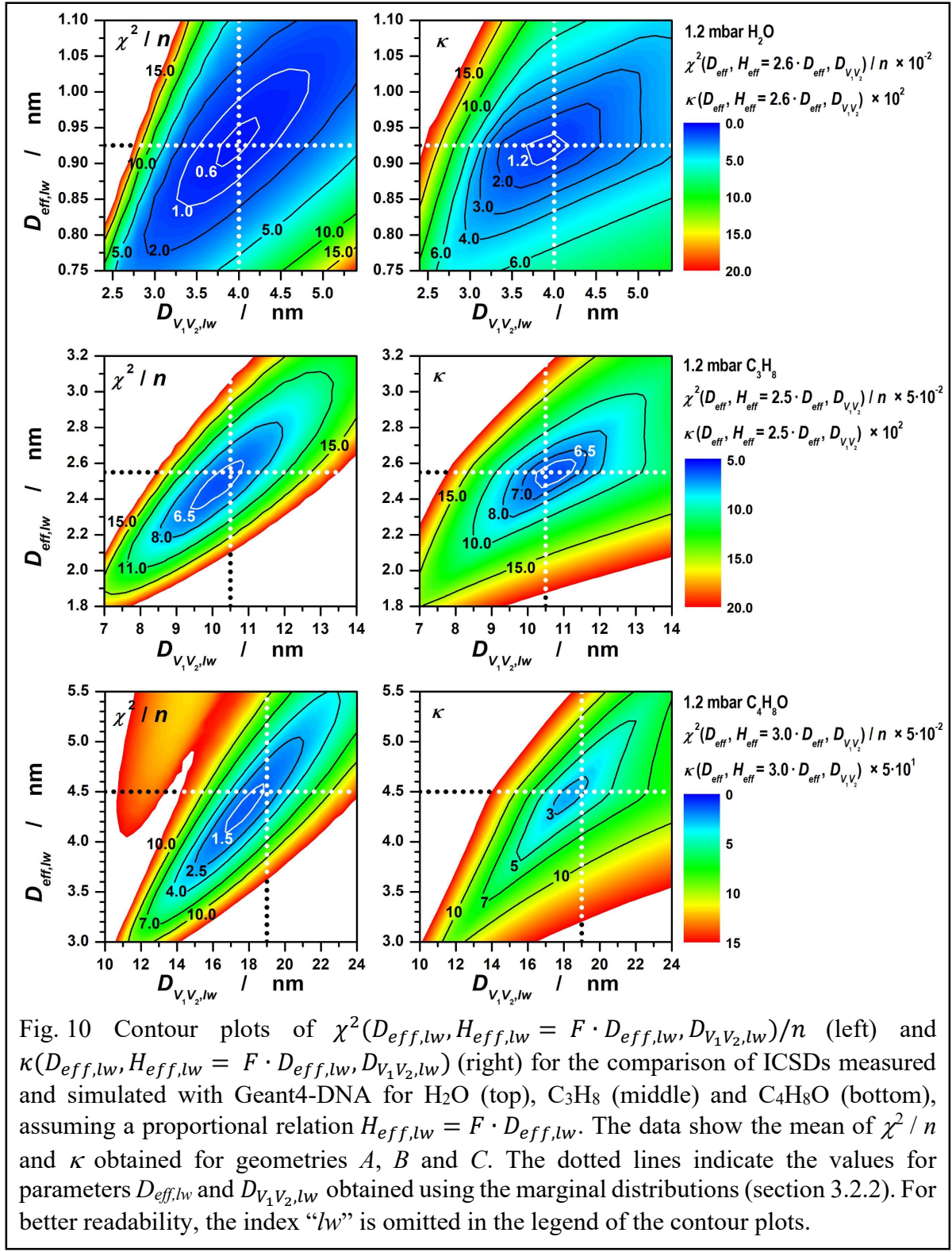

Fig. 10 Contour plots of $\chi^2(D_{eff,lw}, H_{eff,lw} = F \cdot D_{eff,lw}, D_{V_1V_2,lw})/n$ (left) and $\kappa(D_{eff,lw}, H_{eff,lw} = F \cdot D_{eff,lw}, D_{V_1V_2,lw})$ (right) for the comparison of ICSDs measured and simulated with Geant4-DNA for $H_2O$ (top), $C_3H_8$ (middle) and $C_4H_8O$ (bottom), assuming a proportional relation $H_{eff,lw} = F \cdot D_{eff,lw}$. The data show the mean of $\chi^2/n$ and $\kappa$ obtained for geometries *A*, *B* and *C*. The dotted lines indicate the values for parameters $D_{eff,lw}$ and $D_{V_1V_2,lw}$ obtained using the marginal distributions (section 3.2.2). For better readability, the index "*lw*" is omitted in the legend of the contour plots.

The results of the $\chi^2$ test for ICSDs simulated with Geant4-DNA are shown in Table 4 for both, for simulations using the spatial distribution of the extraction efficiency scaled in terms of liquid $H_2O$ and for simulations using the cylinder geometry based on the optimised values for $D_{eff,lw}$, $H_{eff,lw}$ and $D_{V_1V_2,lw}$. Due to the large values of $\chi^2/n$ and $\kappa$ for $C_3H_8$ and $C_4H_8O$ in geometry $B$ (see Table 1), the corresponding data are omitted. Table 4 lists the number of columns $v_{max}$ and rows $\mu_{max}$ of the contingency tables constructed from the distributions $P_{v,\mu}(V_1,V_2,G)$ as well as the resulting ratios $\chi^2/\chi^2_{0.95}$ and $\chi^2/\chi^2_{0.99}$. $\chi^2_{0.95}$ and $\chi^2_{0.99}$ denote the 0.95 and the 0.99 quantiles of the $\chi^2$ distribution with degrees of freedom according to the maximum cluster sizes $v_{max}$ and $\mu_{max}$. The ratios $\chi^2/\chi^2_{0.95}$ and $\chi^2/\chi^2_{0.99}$ indicate the degree of stochastic independence: if the ratio is below unity, the error probability for falsely rejecting the hypothesis that the probability distributions of ICSs in the two sites are statistically independent would exceed 5% or 1%, respectively. In such a case, the distributions would be regarded as stochastic independent, i.e. uncorrelated. For values exceeding unity, the hypothesis of stochastic independence would have to be rejected with an error probability that rapidly decreases with increasing value of $\chi^2$. As can be seen from Table 4, the ratios are strongly dependent on the irradiation geometry. They are maximal for geometry $C$, i.e. with the beam extending over an area covering completely both sites $V_1$ and $V_2$. Except for $H_2O$ in optimised cylinder geometry, the ratios are minimal in geometry $A$, i.e. with the beam passing one of the sites $V_1$ and $V_2$ centrally, and geometry $B$ is in between. In all cases, the ratios are exceeding unity and even for the smallest values found for $H_2O$ in optimised cylinder geometry and geometry $B$, the alpha value for the found value of $\chi^2$ is below $5\times10^{-3}$. Hence, in all cases it must be concluded that the ICSDs for the two sites are correlated. Judging from the ratios listed in Table 4, the degree of correlation seems to be related to the irradiation geometry while no clear dependence on target size can be established.

|  |  | scaled spatial distribution of extraction efficiency | | | optimised cylinder geometry | | |
|---|---|---|---|---|---|---|---|
| Geometry (Fig. 2) | | $A_1$ | B | C | A | B | C |
| $H_2O$ | $v_{max}$ | 9 | 4 | 5 | 8 | 4 | 6 |
| | $\mu_{max}$ | 5 | 5 | 5 | 5 | 5 | 5 |
| | $\chi^2/\chi^2_{0.95}$ | 3.60 | 15.2 | 1324 | 7.08 | 1.31 | 548 |
| | $\chi^2/\chi^2_{0.99}$ | 3.11 | 12.2 | 1088 | 6.05 | 1.05 | 457 |
| $C_3H_8$ | $v_{max}$ | 20 | | 9 | 16 | 7 | 7 |
| | $\mu_{max}$ | 7 | | 9 | 7 | 7 | 7 |
| | $\chi^2/\chi^2_{0.95}$ | 1.57 | | 1444 | 2.20 | 8.22 | 641 |
| | $\chi^2/\chi^2_{0.99}$ | 1.45 | | 1296 | 2.01 | 7.16 | 558 |
| $C_4H_8O$ | $v_{max}$ | 33 | | 12 | 34 | 13 | 15 |
| | $\mu_{max}$ | 7 | | 12 | 10 | 14 | 16 |
| | $\chi^2/\chi^2_{0.95}$ | 3.87 | | 808 | 1.56 | 20.8 | 475 |
| | $\chi^2/\chi^2_{0.99}$ | 3.62 | | 745 | 1.48 | 19.4 | 447 |

Table 4. Number of columns $v_{max}$ and rows $\mu_{max}$ of the contingency table of the distribution $P_{v,\mu}(V_1,V_2,G)$ and the resulting ratios of $\chi^2/\chi^2_{0.95}$ and $\chi^2/\chi^2_{0.99}$ for simulations using Geant4-DNA with geometries $A$ ($A_1$), $B$ and $C$ for the three target gases $H_2O$, $C_3H_8$ and $C_4H_8O$. Due to the large values of $\chi^2/n$ and $\kappa$ for $C_3H_8$ and $C_4H_8O$ in geometry $B$, the corresponding data are omitted.

The results of the $\chi^2$ test for ICSDs simulated with PTra are shown in Table 5 for simulations in 1.2 mbar $C_3H_8$ using the spatial distribution of the extraction efficiency. As for Geant4-DNA, the ratios $\chi^2/\chi^2_{0.95}$ and $\chi^2/\chi^2_{0.99}$ are minimal in geometry $A$. However, different from the simulations with Geant4-DNA, the ratios are of similar magnitude for geometries $B$ and $C$ for simulations with PTra.

| Geometry (Fig. 2) | | $A_1$ | $B$ | $C$ |
|---|---|---|---|---|
| $C_3H_8$ | $\nu_{max}$ | 17 | 7 | 7 |
| | $\mu_{max}$ | 6 | 7 | 8 |
| | $\chi^2/\chi^2_{0.95}$ | 2.21 | 1855 | 1484 |
| | $\chi^2/\chi^2_{0.99}$ | 2.01 | 1614 | 1302 |

Table 5. Number of columns $\nu_{max}$ and rows $\mu_{max}$ of the contingency table of the distribution $P_{\nu,\mu}(V_1,V_2,G)$ and the resulting ratios of $\chi^2/\chi^2_{0.95}$ and $\chi^2/\chi^2_{0.99}$ for simulations using PTra with geometries $A_1$, $B$ and $C$ for 1.2 mbar $C_3H_8$.

Thus, both types of simulations with Geant4-DNA, i.e., using the spatial distribution of the extraction efficiency scaled to liquid $H_2O$ and using the optimised parameters for cylindrical target volumes, as well as the simulations with PTra, confirm the results found in the measurements of correlated production of ionisations in separated target volumes presented in (Hilgers and Rabus, 2020).

The dependence of the correlation of ionisation clusters in the two sites on irradiation geometry and target gas is due to a complex interplay of direct ionisations by the primary ion and by secondary electrons and the electron range in the different target gases and is discussed in detail in (Hilgers and Rabus, 2020). In geometry $A$ the alpha particles pass site $V_1$ centrally (in the equivalent geometry they pass site $V_2$ centrally). The ionisations in the site, which is passed centrally, are due to alpha particles and secondary electrons, and exclusively due to secondary electrons in the site passed by. Correlations are due to the secondary electrons in the site passed by. In geometry $B$ the alpha particles pass through the centre point between the centres of $V_1$ and $V_2$. Ionisations are exclusively due to secondary electrons. Correlations in the ionisations in the two sites can only occur, if at least two secondary electron tracks are created with each of them passing one of the opposite sites. In geometry $C$ the primary beam extends over an area that completely covers both sites $V_1$ and $V_2$. Since most of the alpha particle trajectories pass outside the sites or inside but far from the centre, ionisations due to primaries contribute only sparsely. Therefore, correlations are exclusively due to ionisations created by secondary electrons, which pass through the sites, similar to geometry $B$.

### 3.5. Correlations of ICSDs in other target cylinder arrangements

Besides ICSDs obtained for geometries with two spatially separated target cylinders stacked above each other, there are also other target geometries of relevance, in particular a geometry with the target cylinders of identical size with their central axis' aligned parallel to each other and the cylinder centres aligned in line with the centre of the primary beam, as shown in Fig. 11. However, this geometry is not experimentally accessible with the present setup of the nanodosimeter device and therefore can only be assessed by simulations.

Simulations for this arrangement of two cylindrical target sites were carried out with Geant4-DNA using the optimised parameters $D_{eff,lw}$ and $H_{eff,lw}$ from Table 2 in an irradiation geometry with a primary needle beam (denoted geometry $A^*$) and a second irradiation geometry (denoted geometry $C^*$) with a rectangular beam of the same width and half of the height of the beam used in the simulation for geometry $C$ with $D_{eff,lw}$ and $H_{eff,lw}$. The distance between the two cylinders' axes is 6 nm, i.e. the kinetic energy of the primaries passing the cylinders is practically identical.

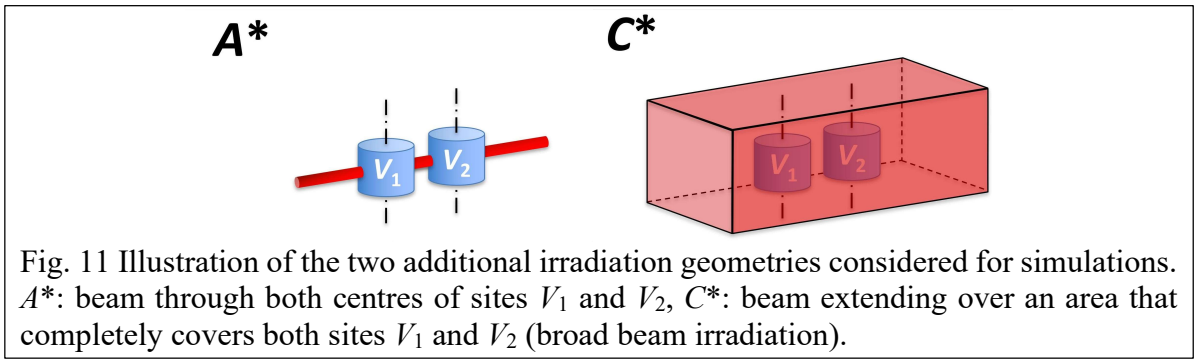

Fig. 11 Illustration of the two additional irradiation geometries considered for simulations. $A^*$: beam through both centres of sites $V_1$ and $V_2$, $C^*$: beam extending over an area that completely covers both sites $V_1$ and $V_2$ (broad beam irradiation).

The resulting bivariate ICSDs are shown in Fig. 12 for geometry $A^*$ in the left column and $C^*$ in the right column. For geometry $A^*$, the bivariate ICSDs have a mirror symmetry with respect to the diagonal $\mu = \nu$, as can be seen clearer in the corresponding contour plots of the simulations for geometry $A^*$ shown in Fig. 13. This implies marginal ICSDs of identical shape and extension for geometry $A^*$. The reason is that the primary alpha particles pass both target sites centrally, in contrast to geometry $A_1$ ($A_2$) where the primaries pass one target site centrally and pass by the second site.

For geometry $C^*$ the differences to geometry $C$ are less obvious. However, closer inspection reveals that the ICSDs of the two geometries differ in two aspects: (i) the margins at cluster sizes $\mu = 0$ and $\nu = 0$ with significantly enhanced relative frequency clearly visible in ICSDs of geometry $C$ are no longer existing in ICSDs of geometry $C^*$ and (ii) in ICSDs of geometry $C^*$ a shoulder appears at cluster size $\mu = \nu \cong 2$ for $D_{eff,lw} = 0.925$ nm ($\mu = \nu \cong 6$ for $D_{eff,lw} = 2.55$ nm) ($\mu = \nu \cong 12$ for $D_{eff,lw} = 4.5$ nm) which does not exist in ICSDs of geometry $C$. Both, the margins in ICSDs of geometry $C$ and the shoulders in ICSDs of geometry $C^*$, represent the contributions of the primary alphas passing the target cylinders through or close to the central axis, with the difference that in geometry $C$ only one target cylinder is passed through or close to the axis by a single primary, whereas in geometry $C^*$ both target cylinders are passed through or close to the cylinder axis by the same primary.

The ICSDs simulated previously for cylindrical target volumes in geometries $B$ and $C$ are symmetrical with respect to the line $\mu = \nu$ due to the symmetric arrangement of the target sites with respect to the primary beam. For geometries $A^*$ and $C^*$ to show the same kind of symmetry requires, that the primary particle's kinetic energy is identical in both target sites. This is the case in the present simulations for geometries $A^*$ and $C^*$ due to the small distance between the two target sites. In the simulations for geometries $B$ and $C$ this requirement is fulfilled automatically due to the arrangement of the target sites relative to the primary beam. For geometries $A^*$ and $C^*$ shifts of the ICSDs from the line of symmetry $\mu = \nu$ will appear for distances between the target sites leading to a sufficient change of the primary particle's kinetic energy (and with it the cross section for ionisation), with the amount of the shift depending on the difference in the cross sections for ionisation.

The change of nanogeometry from geometry $A$ to $A^*$ and from geometry $C$ to $C^*$ significantly affects the degree of correlations in the production of ionisations in separated target sites. The results of the $\chi^2$ test for ICSDs simulated with Geant4-DNA are shown in Table 6 for simulations using the optimised parameter values of the cylindrical target volumes for geometries $A$, $A^*$, $C$ and $C^*$. For geometry $A^*$ the ratios of $\chi^2/\chi^2_{0.95}$ and $\chi^2/\chi^2_{0.99}$ increase with increasing cylinder diameter, indicating an increasing degree of correlation, and are substantially larger than the ratios for geometry $A$ except for the smallest diameter, i.e. for larger cylinders the degree of correlation is larger in geometry $A^*$ than in geometry $A$. For geometry $C^*$ the ratios of $\chi^2/\chi^2_{0.95}$ and $\chi^2/\chi^2_{0.99}$ decrease with increasing cylinder diameter, indicating a decreasing degree of correlation, and are for smaller diameters substantially larger than the ratios for geometry $C$. In general, the degree of correlation in the production of ionisations in separated target sites tends to be more pronounced in geometries $A^*$ and $C^*$ than in the corresponding geometries $A$ and $C$.

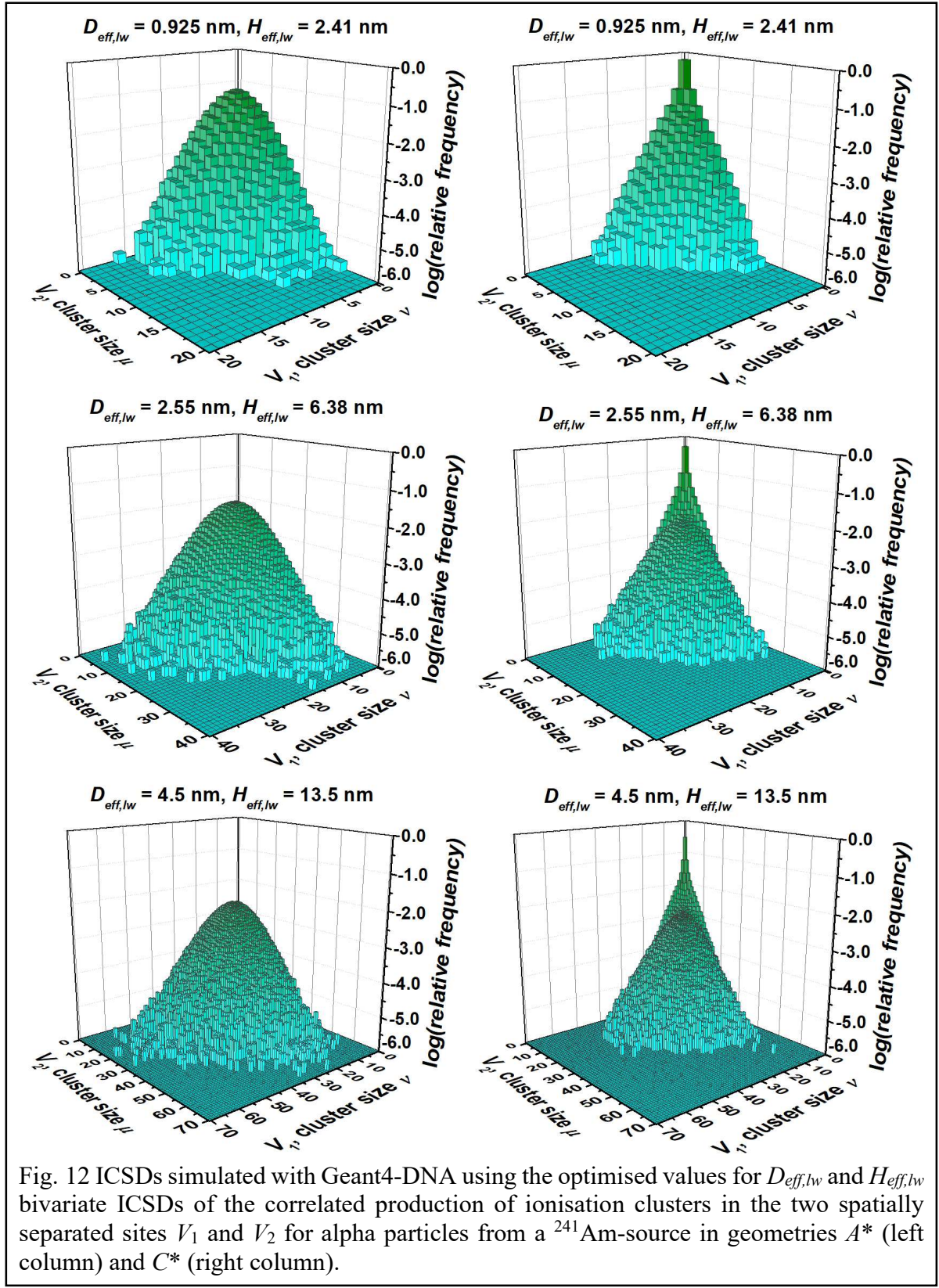

Fig. 12 ICSDs simulated with Geant4-DNA using the optimised values for $D_{eff,lw}$ and $H_{eff,lw}$ bivariate ICSDs of the correlated production of ionisation clusters in the two spatially separated sites $V_1$ and $V_2$ for alpha particles from a $^{241}$Am-source in geometries $A^*$ (left column) and $C^*$ (right column).

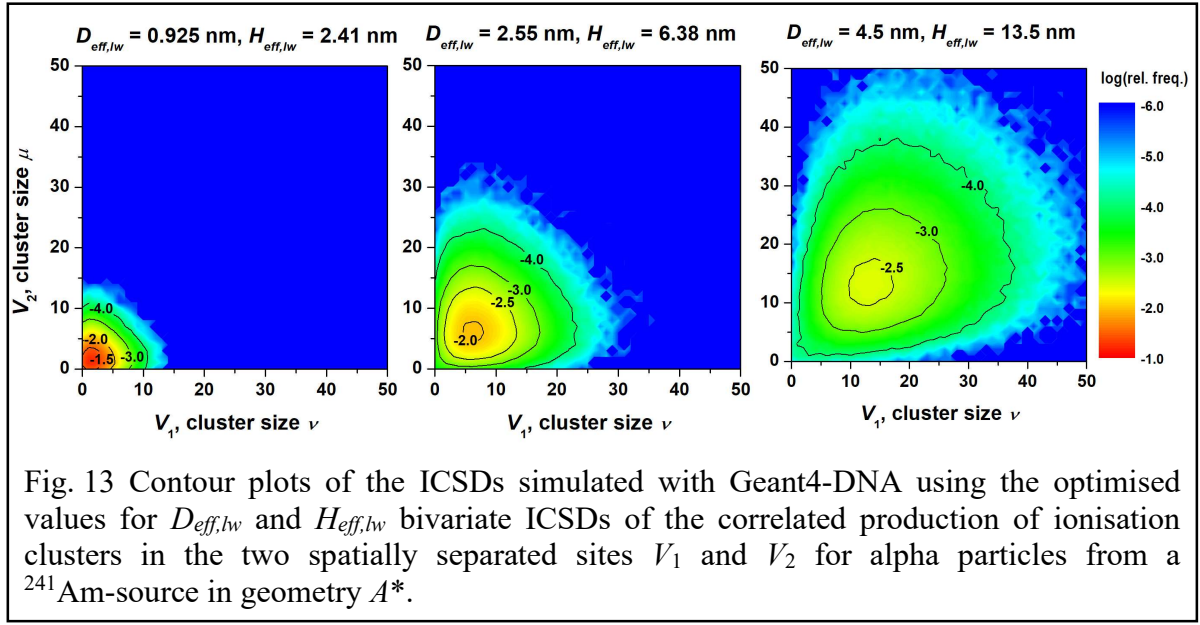

Fig. 13 Contour plots of the ICSDs simulated with Geant4-DNA using the optimised values for $D_{eff,lw}$ and $H_{eff,lw}$ bivariate ICSDs of the correlated production of ionisation clusters in the two spatially separated sites $V_1$ and $V_2$ for alpha particles from a $^{241}$Am-source in geometry $A^*$.

In previous investigations the cumulative probability $F_k(V,G)$ (see eq. (4) in (Hilgers and Rabus, 2020)) for $k = 2$ was assumed to show a similar dependence on radiation quality as the yields of double strand breaks of the DNA for broad-beam irradiation and a site of cylindrical shape with 2.3 nm diameter and 3.4 nm height (Grosswendt, 2004; 2005; 2007). More recently, $F_k(V,G)$ were found to correlate with the cross section of inactivation of different cell lines in the sense that both show a similar dependence on radiation quality (Conte et al., 2017; 2018). In these investigations, the geometry was central passage of a cylindrical site with equal diameter and height of 1 nm for $k$ = 2 and of 1.5 nm for $k$ = 3. In view of these findings also the corresponding quantities $F_{k,k}(V_1,V_2,G)$ (see eq. (11) in (Hilgers and Rabus, 2020)) for the correlated production of $v, \mu \geq k$ ionisations in two spatially separated sites $V_1$ and $V_2$ may be expected to be relevant for radiation effects on the DNA for proper dimensions of the sites. So, for suitable site dimensions, $F_{2,2}(V_1,V_2,G)$ might indicate the probability for occurrence of double strand breaks in two DNA segments in close vicinity and $F_{3,3}(V_1,V_2,G)$ might indicate the occurrence of complex double strand breaks in two DNA segments in close vicinity. Hence, both quantities might be related to a possible loss of DNA loops due to correlated induction of double strand breaks in two neighbouring DNA segments.

| Geometry | | $A$ | $A^*$ | $C$ | $C^*$ |
|---|---|---|---|---|---|
| $H_2O$, $D_{eff,lw}$ = 0.925 nm | $\nu_{max}$ | 8 | 11 | 6 | 10 |
| | $\mu_{max}$ | 5 | 11 | 5 | 9 |
| | $\chi^2/\chi^2_{0.95}$ | 7.08 | 2.10 | 548 | 2510 |
| | $\chi^2/\chi^2_{0.99}$ | 6.05 | 1.92 | 457 | 2266 |
| $C_3H_8$, $D_{eff,lw}$ = 2.55 nm | $\nu_{max}$ | 16 | 25 | 7 | 21 |
| | $\mu_{max}$ | 7 | 26 | 7 | 22 |
| | $\chi^2/\chi^2_{0.95}$ | 2.20 | 4.14 | 641 | 1000 |
| | $\chi^2/\chi^2_{0.99}$ | 2.01 | 3.98 | 558 | 957 |
| $C_4H_8O$, $D_{eff,lw}$ = 4.5 nm | $\nu_{max}$ | 34 | 41 | 15 | 37 |
| | $\mu_{max}$ | 10 | 43 | 16 | 35 |
| | $\chi^2/\chi^2_{0.95}$ | 1.56 | 7.53 | 475 | 473 |
| | $\chi^2/\chi^2_{0.99}$ | 1.48 | 7.36 | 447 | 462 |

Table 6. Number of columns $\nu_{max}$ and rows $\mu_{max}$ of the contingency table of the distribution $P_{\nu,\mu}(V_1,V_2,G)$ and the resulting ratios of $\chi^2/\chi^2_{0.95}$ and $\chi^2/\chi^2_{0.99}$ for simulations using Geant4-DNA with geometries $A$, $A^*$, $C$ and $C^*$ for the three target gases and the three sets of optimised parameter values of the target cylinders, respectively.

| Geometry | | A | A* | C | C* |
|---|---|---|---|---|---|
| $H_2O$, $D_{eff,lw}$ = 0.925 nm | $F_{2,2}(V_1,V_2)$ | 0.0015 | 0.3606 | 0.0005 | 0.0626 |
| | $F_{3,3}(V_1,V_2)$ | 0.0003 | 0.1339 | 0.0001 | 0.0195 |
| $C_3H_8$, $D_{eff,lw}$ = 2.55 nm | $F_{2,2}(V_1,V_2)$ | 0.0080 | 0.9482 | 0.0030 | 0.1726 |
| | $F_{3,3}(V_1,V_2)$ | 0.0044 | 0.8699 | 0.0013 | 0.1357 |
| $C_4H_8O$, $D_{eff,lw}$ = 4.5 nm | $F_{2,2}(V_1,V_2)$ | 0.0129 | 0.9964 | 0.0089 | 0.3506 |
| | $F_{3,3}(V_1,V_2)$ | 0.0092 | 0.9907 | 0.0054 | 0.3090 |

Table 7. Cumulative probabilities $F_{2,2}(V_1,V_2,G)$ and $F_{3,3}(V_1,V_2,G)$ for simulations using Geant4-DNA with geometries A, A*, C and C* for the three target gases and the three sets of optimised parameter values of the cylindrical target volumes, respectively.

Table 7 shows the cumulative probabilities $F_{2,2}(V_1,V_2,G)$ and $F_{3,3}(V_1,V_2,G)$ for simulations using Geant4-DNA with geometries A, A*, C and C* for the three target gases and the three sets of optimised parameter values of the cylindrical target volumes, respectively. Both, $F_{2,2}(V_1,V_2,G)$ and $F_{3,3}(V_1,V_2,G)$, increase with increasing target size for all geometries. For the geometries with the target cylinder centres aligned in line with the centre of the primary beam, i.e. geometries A* and C*, the values found for both, $F_{2,2}(V_1,V_2,G)$ and $F_{3,3}(V_1,V_2,G)$, are substantially larger (up to two orders of magnitude) than for the geometries A and C, where the target cylinders are stacked above each other, with the values for the narrow beam geometry A* are 3 – 6 times larger than for the broad beam geometry C*. This indicates that narrow beam geometries with target sites located directly on the trajectory of the primary particle are most effective in the correlated production of ionisation clusters. However, also corresponding broad beam geometries, which are the most relevant with respect to radiation therapy, with target sites located along the propagation direction of the primary particles are of similar effectiveness.

Table 8 shows the correlation coefficients $R(V_1,V_2,G)$ and $R_{k,k}(V_1,V_2,G)$ of the bivariate ICSDs and of the cumulative probabilities $F_{k,k}(V_1,V_2,G)$ for $k$ = 2, 3, respectively, for simulations using Geant4-DNA with geometries A, A*, C and C* for the three target gases and the three sets of optimised parameter values of the cylindrical target volumes. All three correlation coefficients confirm the findings from the ratios of $\chi^2/\chi^2_{0.95}$ and $\chi^2/\chi^2_{0.99}$ of the $\chi^2$ independence test.

Both geometries, A and A*, show only a very small degree of correlation, i.e. the production of ionisation clusters in the two target sites is only weakly correlated and the resulting bivariate ICSDs and their cumulative probabilities are dominated by coincidences rather than correlations. For geometry $A^*$ the degree of correlation increases slightly with the target size, whereas for geometry $A$ no dependency on the target size is visible.

| Geometry | | A | A* | C | C* |
|---|---|---|---|---|---|
| $H_2O$, $D_{eff,lw}$ = 0.925 nm | $R(V_1,V_2,G)$ | 0.0152 | 0.0038 | -0.0941 | 0.4228 |
| | $R_{2,2}(V_1,V_2,G)$ | 0.0061 | 0.0042 | -0.0683 | 0.3638 |
| | $R_{3,3}(V_1,V_2,G)$ | 0.0036 | 0.0021 | -0.0349 | 0.2226 |
| $C_3H_8$, $D_{eff,lw}$ = 2.55 nm | $R(V_1,V_2,G)$ | 0.0093 | 0.0147 | -0.1095 | 0.6133 |
| | $R_{2,2}(V_1,V_2,G)$ | 0.0039 | 0.0258 | -0.1287 | 0.5518 |
| | $R_{3,3}(V_1,V_2,G)$ | 0.0042 | 0.0222 | -0.1023 | 0.5894 |
| $C_4H_8O$, $D_{eff,lw}$ = 4.5 nm | $R(V_1,V_2,G)$ | 0.0111 | 0.0470 | -0.1843 | 0.6821 |
| | $R_{2,2}(V_1,V_2,G)$ | 0.0025 | 0.0322 | -0.2658 | 0.5431 |
| | $R_{3,3}(V_1,V_2,G)$ | 0.0026 | 0.0636 | -0.2304 | 0.5978 |

Table 8. Correlation coefficients $R(V_1,V_2,G)$ and $R_{k,k}(V_1,V_2,G)$ of the bivariate ICSDs and of the cumulative probabilities $F_{k,k}(V_1,V_2,G)$ for $k = 2, 3$, respectively, for simulations using Geant4-DNA with geometries $A$, $A^*$, $C$ and $C^*$ for the three target gases and the three sets of optimised parameter values of the cylindrical target volumes.

For geometries $C$ and $C^*$ the degree of correlation is clearly enhanced compared to geometries $A$ and $A^*$, with geometry $C^*$ showing a more pronounced degree of correlation than geometry $C$. For both geometries the degree of correlation increases with target size as can be seen in the absolute values of the correlation coefficients. For geometry $C$ the numerical values of the correlation coefficients indicate a moderate correlation and are negative, i.e. an increase in cluster size in one target site tends to lead to a decrease in cluster size in the other target site and vice versa. The corresponding bivariate ICSDs and their cumulative probabilities are to some extent determined by correlations but are still significantly affected by coincidences. For geometry $C^*$ the correlation coefficients indicate a pronounced correlation. As they are positive, an increase in cluster size in one target site is connected to an increase in cluster size in the other target site and vice versa. The corresponding bivariate ICSDs and their cumulative probabilities are substantially determined by correlations and to some extent influenced by coincidences.

## 5. Summary and conclusion

A dedicated investigation of the size of two spatially separated target sites involved in the correlated production of ionisations of a nanodosimeter device has been carried out, with particular focus on the equivalent target size in terms of liquid $H_2O$.

The method, which has been developed in (Hilgers et al., 2021) to determine the dimensions of a single simulated nanometric target volume in liquid $H_2O$ of cylindrical shape, as often used in approaches to model radiation effects to DNA, has been successfully extended to the determination of two target volumes of cylindrical shape and the determination of the distance separating the two target volumes, both in terms of liquid $H_2O$. Two different approaches were used to determine the parameters of the simple geometry of two stacked cylinders. In the first approach the parameters of the geometry were optimised consecutively. In the second approach the parameters describing the geometry were optimised simultaneously. The values found for the parameters describing the geometry with the two different approaches agree within their estimated uncertainties. Therefore, both approaches can be regarded as equivalent procedures. Simulations with nanometric targets of dimensions determined with this method agree very well with the corresponding measurements.

For the considered radiation quality, and for the irradiation geometry with the primary beam passing one of the target sites centrally and passing by the other and for the irradiation geometry with a broad beam covering both target sites, the dimensional scaling of the spatial distribution of the extraction efficiency is appropriate to estimate the target volume in liquid $H_2O$. In the geometry considered, where the ionisations in the target sites are practically exclusively due to ionisations of secondary electrons, since the primary alpha particles pass through the centre point between the centres of the target sites, large values for the reduced $\chi^2$ and $\kappa$ are found, indicating a poorer agreement. The scaling procedure described in (Grosswendt, 2006) is only valid for ionisations due to the primary alpha particles and not for the secondary electrons in the penumbra of the primary particle track. Consequently, the scaling procedure is not applicable in geometries where ionisations are predominantly created by secondary electrons.

The simulated bivariate distributions $P_{\nu,\mu}(V_1,V_2,G)$ were tested for stochastic independence by application of a $\chi^2$ independence test. The degree of correlation has been found to be determined mostly by the irradiation geometry. A correspondence between the degree of correlation and the size of the target volume, and thus a dependence on the target gas, does not seem to exist. These findings for the simulations confirm the results found in the measurements of correlated production of ionisations in separated target volumes presented in (Hilgers and Rabus, 2020). Interpreting the dependence of the correlation on irradiation geometry and target gas requires consideration of a complex interplay of direct ionisations by the primary ion and by secondary electrons and the electron range in the different target gases.

In addition to the ICSDs for geometries simulated by the nanodosimeter device, geometries with the spatially separated target cylinders aligned in line with the trajectories of the primary particles, i.e. in the propagation direction of the primary beam, were investigated. Due to the setup of the nanodosimeter these geometries are not experimentally accessible but only by simulation. These geometries are found to be much more effective in the simultaneous production of ionisation clusters in the two target sites: the cumulative probabilities, $F_{k,k}(V_1,V_2,G)$ for $k = 2, 3$, are up to two orders of magnitude larger than for the geometries, where the target cylinders are stacked above each other.

However, not in all geometries the bivariate ICSDs and their cumulative probabilities are governed by correlations. There are geometries where the production of ionisation clusters in the two target sites is only weakly correlated and the resulting bivariate ICSDs and their cumulative probabilities are dominated by coincidences rather than correlations. On the other hand, geometries do exist with pronounced degree of correlation in the production of ionisations in two target sites where the corresponding bivariate ICSDs and their cumulative probabilities are dominantly determined by correlations. Moreover, geometries can be found with negative correlation coefficients, where an increase in cluster size in one target site leads to a decrease in cluster size in the other target site and vice versa.

For suitable site dimensions, $F_{2,2}(V_1,V_2,G)$ might indicate the probability for simultaneous occurrence of double strand breaks in two DNA segments and $F_{3,3}(V_1,V_2,G)$ might indicate the simultaneous occurrence of complex double strand breaks in two DNA segments. Hence, both quantities might be related to a possible loss of DNA loops due to correlated induction of double strand breaks in two spatially separated DNA segments. Taking into account the large number of DNA segments in a cell nucleus, is seems highly probable, that two of more segments of proper alignment are found on a primary particle's trajectory, especially in broad beam geometries, which are the most relevant in radiation therapy. Therefore, it might be useful to consider the correlated production of ionisations in the description of radiation induced damage to DNA, in particular for densely ionising particles, e.g., ions used in hadron radiation therapy, and for primary particle kinetic energies in the Bragg peak region, where the amount of ionisations is maximal. However, at present, a dedicated investigation of this aspect of radiation damage to DNA seems realistic only by simulation until suitable measuring devices will have been developed.

## Acknowledgement

The authors gratefully acknowledge the developers of the nanodosimeter from the Weizmann Institute of Science, Rehovot, Israel, for transferring the device as described in (Garty et al., 2002) to PTB for further use. The authors also express their gratitude to B. Lambertsen and A. Pausewang for their invaluable contributions to preparation and carrying out the measurements and their assistance in data processing. H. Nettelbeck is acknowledged for proof-reading the manuscript.